\documentclass[aps,prb,amsmath,amssymb,superscriptaddress,notitlepage,twocolumn]{revtex4-2}
\usepackage{graphicx}
\usepackage{bm}
\usepackage{amsmath}
\usepackage{amssymb}
\usepackage{mathrsfs}
\usepackage{latexsym}
\usepackage{amsbsy}
\usepackage{array}
\usepackage{amssymb}
\usepackage{bm}
\usepackage{color}
\usepackage{ulem}
\usepackage[colorlinks, urlcolor=blue, citecolor=blue]{hyperref}
\usepackage{tikz-feynman}
\usepackage{makecell}
\usepackage{siunitx}

\let \oldbm \bm
\renewcommand{\vec}[1]{\oldbm{#1}}

\def\bk{{\vec k}}

\def\bxi{{\vec \xi}}
\def\bL{{\vec L}}
\def\bu{{\vec u}}

\def\ba{{\vec a}}
\def\bb{{\vec b}}

\def\bq{{\vec q}}

\def\bO{{\vec O}}
\def\bG{{\vec G}}

\def\bx{{\bf x}}

\def\bm{{\vec m}}

\def\br{{\vec r}}

\def\bphi{{\boldsymbol \phi}}

\def\bdelta{{\boldsymbol \delta}}
\def\bepsilon{{\boldsymbol \epsilon}}

\def\tr{\mathop{\mathrm{tr}}}

\def\L{\mathcal{L}}
\def\E{\mathcal{E}}

\def\C{\mathcal{C}}

\def\H{\mathcal{H}}

\def\M{\mathcal{M}}
\def\B{\mathcal{B}}

\def\A{\mathcal{A}}

\def\R{\mathcal{R}}

\begin{document}
\title{Symmetry Origin of Lattice Vibration Modes in Twisted Multilayer Graphene: Phasons vs Moir\'e Phonons}
\author{Qiang Gao}
\affiliation{Department of Physics, The University of Texas at Austin, TX 78712, USA}
\author{Eslam Khalaf}
\affiliation{Department of Physics, The University of Texas at Austin, TX 78712, USA}
\affiliation{Department of Physics, Harvard University, Cambridge, MA 02138, USA}
\begin{abstract}
Lattice dynamics play a crucial role in the physics of moir\'e systems. In twisted bilayer graphene (TBG), it was shown that, in addition to the graphene phonons, there is another set of gapless excitations termed moir\'e Phonons [Phys. Rev. B, 075416, 2019] reflecting the lattice dynamics at the moir\'e superlattice level.  These modes were later suggested to be phasons due to the incommensurate stacking of the two graphene layers [Phys. Rev. B, 155426, 2019]. In this work, we elucidate the equivalence of these two seemingly distinct perspectives by identifying an underlying symmetry, which we dub mismatch symmetry, that exists for any twist angle. For commensurate angles, this is a discrete symmetry whereas, for incommensurate angles, it is equivalent to a continuous phase symmetry giving rise to phason modes. In the small angle limit, such symmetry becomes a continuous local symmetry whose spontaneous breaking gives rise to moir\'e phonons as its Goldstone mode. We derive an effective field theory for these collective modes in TBG in precise agreement with the full model and discuss their different properties. Our analysis is then generalized to twisted multilayer graphene (TMG) where we identify higher-order mismatch and deduce the count of gapless modes including graphene phonons, moir\'e phonons, and phasons.
Especially, we study twisted mirror-symmetric trilayer graphene with an alternating twist angle $\theta$ and find that it can be mapped to a TBG with the re-scaled twist angle $\sqrt{2/3}\theta$, hosting the same moir\'e phonon modes in the even mirror sector with an additional set of gapped modes in the odd sector. Our work presents a systematic study of lattice symmetries in TMG providing insights into its unique lattice dynamics.
\end{abstract}
\maketitle

\section{Introduction}
In recent years, twisted bilayer graphene (TBG) has emerged as a novel tunable platform to study strongly interacting electronic phases~\cite{cao2018unconventional,cao2018correlated,yankowitz2019tuning, Balents2020Review}. The lattice configuration of the TBG plays an important role not only for the electronic structure~\cite{nam2017lattice} but also for the lattice dynamics~\cite{koshino2019moire,ochoa2019moire} which could be relevant to explaining several experimental observations including superconductivity~\cite{wu2018theory,lian2019twisted} and normal state transport behavior~\cite{Sharma2021}. In addition to the acoustic phonons inherited from single layer graphene, TBG hosts a set of moir\'e phonons \cite{koshino2019moire} corresponding to the relative movement of the two layers. The latter can be understood as the phonons of the domain wall network formed at small twist angles when TBG mechanically relaxes.  An insightful work from Ochoa \cite{ochoa2019moire} has argued that these moir\'e phonons are in fact phasons of the incommensurate moire superlattice. This idea has been elaborated in a recent work \cite{ochoa2022degradation} which discussed the implication of twist-angle disorder and anharmonic terms on these modes.

However, it is unclear how to reconcile these two seemingly different perspectives: a phason is a collective mode associated with the (generally non-local) phase shift symmetry of an incommensurate pattern and whose dynamics are generally diffusive whereas a phonon is a true propagating Goldstone mode associated with a spontaneously broken local continuous symmetry. This poses several questions: How can these two excitations be the same? What underlying symmetry is responsible for the emergence of these modes? and what role does lattice commensurability play in determining the nature of the excitation? In addition to these conceptual questions, there are also some quantitative discrepancies between the spectrum computed using the moir\'e phonon perspective \cite{koshino2019moire} and that obtained from the phason effective field theory \cite{ochoa2019moire}. For instance, moir\'e phonons can be understood as the phonons of an effective network model of domain walls at a small angle. This perspective predicts a universal ratio of the velocities of the transverse and longitudinal modes that is equal to $\sqrt{3} \approx 1.7$ independent of the material parameters. On the other hand, the phason effective field theory predicts a material-dependent ratio given by $\approx 1.24$ for the graphene parameters \cite{ochoa2019moire}.

These conceptual distinctions are particularly relevant when considering multilayer systems with multiple twist angles \cite{zhu2020modeling,zhu2020twisted,zhang2021correlated}. For instance, a trilayer system with two relative twists angles $\theta_{21}$ and $\theta_{32}$ will realizes a so-called moir\'e of moir\'e \cite{zhu2020modeling, zhu2020twisted}, whenever $\theta_{21} \neq \theta_{23}$. This is a moir\'e pattern formed from the two moir\'e patterns of layers 1 and 2 and that of layers 2 and 3. Understanding the collective lattice modes in these systems require careful consideration of their quasi-periodic nature and the continuum limit. The special case of alternating twist angle $\theta_{21} = -\theta_{32} = \theta$ \cite{khalaf2019magic} have received considerable interest recently following the discovery of superconductivity in this system around a magic angle of $\theta \approx 1.5^\circ$~\cite{park2021tunable, KimTrilayer}. Although this structure has a single moir\'e superlattice rather than a moir\'e of moir\'e, understanding its lattice dynamics requires still requires careful consideration of the commensurability effects and what happens in the continuum limit.

In this work, we present a general analysis of the underlying symmetries responsible for the soft lattice vibrations in twisted multilayer moir\'e systems. We will show that all possible soft collective modes in the system (apart from the trivial phonons inherited from graphene) are a consequence of a certain symmetry of the stacked heterostructure which we dub the mismatch symmetry. This symmetry reduces to a discrete symmetry for commensurate stacking and to a continuous but non-local symmetry for incommensurate stacking. We show that the latter is equivalent to the phase symmetry known in aperiodic lattices which gives rise to phason modes \cite{lubensky1985hydrodynamics,ZeyherFinger, FingerRice, landry2020effective, widom2008discussion}. In the continuum limit, the mismatch symmetry becomes a local continuous symmetry whose breaking gives rise to propagating Goldstone modes which we identify with the moir\'e phonons. This establishes the equivalence of moir\'e phonons and phasons in the continuum limit where the distinction between commensurate and incommensurate stacking is lost. An important implication of this analysis is these modes are true propagating Goldstone modes within the continuum approximation such that any possible damping is suppressed as small angles. This discussion is generalized to multilayer systems with multiple twist angles where we provide a general rule for the count of propagating moir\'e phonons and diffusive phasons. 

We then formulate a low-energy effective field theory for these modes in TBG, revealing that the moir\'e acoustic phonons are indeed Goldstone modes corresponding to the spontaneous breaking of the mismatch symmetry. The theory has the same form as that derived in Ref.~\cite{ochoa2019moire}, but differs in the values of the parameters making it in full agreement with numerical results \cite{koshino2019moire}. We also show it is reduced to the results of a phenomenological model for a domain wall network at a small angle \cite{koshino2019moire}. We then apply our formalism to alternating twist trilayer graphene (ATTG) whose lattice dynamics decouples into an even and odd mirror sector. We find that, rather remarkably, the dynamics of the even mirror sector at angle $\theta$ map exactly to that of TBG at twist angle $\sqrt{2/3} \theta$. This differs from  the electronic spectrum which maps to TBG at $\theta/\sqrt{2}$ \cite{khalaf2019magic}. On the other hand, we find that the mirror-odd spectrum is fully gapped in full accordance with our formalism. We note that, unlike the electronic mapping, this mapping from ATTG to TBG does not have a straightforward generalization to alternating twist multilayer systems \cite{khalaf2019magic, Ledwith2021tb}.

\section{Symmetries, commensurability, and the continuum limit}
In this section, we discuss the different symmetries of a system composed of subsystems with different periods and how such symmetries behave in the continuum limit. We will start by considering the simplest possible setting of a two-chain model \cite{ChaikinLubensky, RadulescuJanssen, FingerRice} which serves to illustrate the main conceptual ingredients of our analysis before generalizing to twisted bilayer and multilayer graphene systems.

\subsection{Two chain model}
\subsubsection{Symmetries: commensurate vs incommensurate periods}\label{symmInTCM}
\begin{figure}
    \centering
    \includegraphics[width=0.48\textwidth]{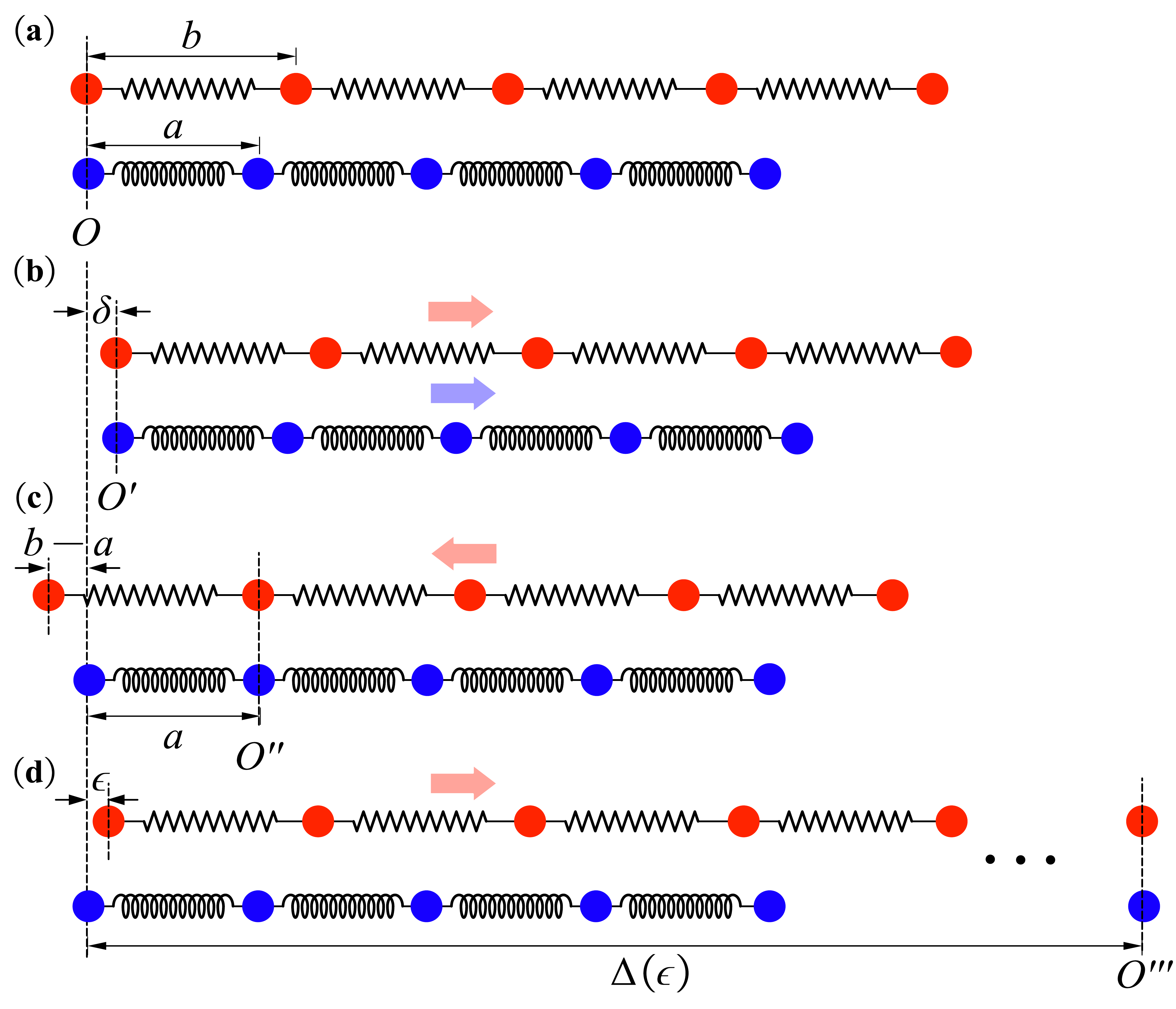}
    \caption{The symmetries in the interacting two-chain model. (a) The initial configuration of the two chains with lattice constants $a$ and $b$. $O$ is the matching point. (b) Two chains move to the right by the same distance $\delta$. The matching point is shifted to point $O'$. (c) The top chain moves to the left by a distance $b-a$ resulting in a shift of the matching point to $O''$. (d) The top chain moves to the right by a distance $\epsilon$ which will effectively shift the matching point to $O'''$. The colored arrows indicate the direction of the movements.}
    \label{fig:two_chain}
\end{figure}
Consider two mono-atomic chains with different lattice spacing, $a$ and $b$ ($a<b$), stacking in parallel, denoted as chain 1 and chain 2, respectively. We will first consider the case where there is no lattice relaxation which corresponds to the limit of vanishing interlayer interaction. We will discuss later how the analysis is generalized for the case of lattice relaxation. We choose to align those two chains such that one lattice point from chain 2 is exactly on the top of one lattice point from chain 1, as shown in Fig.~\ref{fig:two_chain}(a). We call such an exactly aligned point the matching point $O$. Thus, one can immediately see that the position of the matching point $O$ uniquely determines the non-relaxed lattice configuration.

It is obvious that this structure has a continuous translation symmetry corresponding to the spontaneous shifting of the two layers by the same distance  $\delta$ as illustrated in Fig.~\ref{fig:two_chain}(b). This operation shifts the point $O$ to $O'=O+\delta$ and it is the symmetry that gives rise to standard acoustic phonons. The two-chain model has another symmetry. To see this, first, note that shifting the top chain by $b$ does nothing and it mainly amounts to relabelling the lattice sites. On the other hand, shifting the top chain by $a$ shifts the center of the beating pattern from $O$ to $O''=O+a$. This gives rise to a distinct configuration that has the same energy as the original one. Since translation by $b$ does nothing, we can identify the elements of this symmetry group by shifting the top layer by $q_n = n a \mod b$. Such shift generates a shift of the center of the beating pattern $O$ by $n a$ which we dub mismatch symmetry.

If the periods $a$ and $b$ are commensurate, i.e. $\frac{a}{b} \in \mathbb{Q}$, then there exists a minimal $N > 0$ such that $N a \mod b = 0$. Then the symmetry group of mismatch symmetry is isomorphic to $\mathbb{Z}_N$ and its generator can be chosen to be $q_1 = a \mod b$ which shifts the matching point $O$ by a. In Fig.~\ref{fig:two_chain}(c), we show the action of such symmetry. Since we have chosen $a$ and $b$ such that $b-a < a$, the symmetry is implemented as a shift to the left by $b-a$ which corresponds to the shift $q = -b + a \mod b = a \mod b = q_1$ where the matching point shifts from $O$ to $O''=O+a$. Note that for $a = b$, $a \mod b = 0$ and the symmetry group is trivial.

On the other hand, if the periods $a$ and $b$ are incommensurate, i.e. $\frac{a}{b} \notin \mathbb{Q}$, the elements $q_n = n a \mod b$ provide a dense covering of the interval $[0,b)$. Thus, for any shift of the top chain by $\epsilon$, there will always be two integer numbers, $N_1$ and $N_2$, such that $N_1 a - N_2 b$ is arbitrarily close to $\epsilon$. This will provide the symmetry action depicted in Fig.~\ref{fig:two_chain}(d) where the top chain is shifted by $\epsilon$ resulting in a shift in the beating pattern by $\Delta(\epsilon) = N_1 a$. In this case, the symmetry is arbitrarily close to a continuous symmetry but the resulting transformation of the matching point is highly non-local. By locality, we mean that a small shift in the lattice should result in a small shift in the matching point. We would like to point out that this is actually the phase symmetry which gives rise to the phason excitation in incommensurate lattices. In Appendix~\ref{equivalenceRelation}, we discuss in detail that this phase symmetry is equivalent to the mismatch symmetry. The unusual property of the phase symmetry where an infinitesimal shift $\epsilon $ could lead to an infinite $\Delta(\epsilon)$, i.e. $\epsilon\to 0 \Rightarrow\Delta(\epsilon)\to \infty$ indicates that it can not be infinitesimally generated in the underlying dimensions and extra dimensions have to be invoked~\cite{widom2008discussion}. Notice that this property implies that although the Lagrangian is invariant under the infinitesimal shift $\epsilon$, the Lagrangian density does not change by a total derivative under this shift due to the singular dependence of $\Delta(\epsilon)$. This is the underlying reason for the absence of a Noether current for the phase symmetry \cite{landry2020effective}. It implies that phasons are not true Goldstone modes and their dispersion can acquire a diffusive term from anharmonic effects \cite{ZeyherFinger, FingerRice}.

\subsubsection{Continuum limit}
Let us now see what happens when $b \approx a$ such that $1 - \frac{a}{b} = \delta \ll 1$. In this case, there is a new emergent length scale (which we will also call moir\'e scale) given by $L_M = a / \delta \gg a$~\footnote{The typical length is adopted from Ref.~\cite{nam2017lattice} for a commensurate stacking of two chains, where $a=L_M/N$ and $b=L_M/(N-1)$ for some large $N\in \mathbb{Z}$. For general stacking order where $\delta=1-\frac{a}{b}\ll 1$, it can be generalized to $L_M=a/\delta$, which is a good approximation for the typical length scale in analogy to the moir\'e length in TBG. Notice that $L_M$ needs not to be equal to the exact period of the commensurate pattern where $a/b\in \mathbb{Q}$.}. For physics at that scale, we can take the continuum limit $a(\approx b)/L_M \rightarrow 0$. We will now argue that (i) the distinction between the commensurate and incommensurate cases is gone and (ii) the mismatch symmetry becomes a continuous local symmetry. Point (i) can be seen by noting that the main distinction between commensurate and incommensurate cases is that the equation $N a \!\!\mod b = 0$ has a finite solution $N$ in the former but no finite solution in the latter. In the continuum limit, we take $N_\delta = \lfloor \frac{1}{\delta} \rfloor = \frac{1}{\delta} - \{\frac{1}{\delta} \} = \frac{b}{b-a} - \{\frac{1}{\delta} \}$ (where $\{ x \}$ denotes the fractional part of $x$) which gives
\begin{equation}
    N_\delta a \!\!\!\!\mod b = N_\delta (a - b) \!\!\!\!\mod b = -\{\frac{1}{\delta} \} b \delta \!\!\!\!\mod b \sim O(b \delta)
\end{equation}
Thus, $N_\delta$ provides a solution to the equation $N a \!\!\mod b = 0$ up to corrections of order $\delta = \frac{a}{L_M}$ which is taken to zero in the continuum limit. Point (ii) is shown by noting that the mismatch symmetry group is generated by the element $q_1 = a \mod b = b \delta \mod b$ and it is isomorphic to $\mathbb{Z}_{N_\delta}$ which provides a dense covering of the interval $[0,b)$ in the limit $\delta \rightarrow 0$. The crucial distinction here from the incommensurate case is that an infinitesimal shift on the lattice scale $b \delta$ generates an infinitesimal shift of the matching point $O$ on the moir\'e scale $\Delta(b \delta) = a = L_M \delta $. Thus, the mismatch symmetry becomes a local continuous symmetry in the continuum limit.

\subsubsection{Lattice relaxation}
So far, we have ignored the effects of lattice relaxations which means that we assumed the position of atoms is given by the lattice points in each chain $\phi_n^{(1)} = n a$ and $\phi_n^{(2)} = n b$. Lattice relaxation modifies the atomic positions to optimize the inter-chain interaction as
\begin{equation}
    \phi_n^{(l)} = n a^{(l)} + u_n^{(l)}, \qquad a^{(1)} = a, \quad a^{(2)} = b,
\end{equation}
where $u_n^{(l)} $ is the displacement of lattice point $n$ in $l^{\text{th}}$ chain.
The energy consists of an intra-chain part which depends on $\phi_n^{(l)} - \phi_m^{(l)}$ for $l = 1,2$ and an inter-chain part depending $\phi_n^{(1)} - \phi_m^{(2)}$. 

In the continuum limit, we can write $x = n b$ such that 
\begin{equation}
    \phi^{(1)}(x) = x (1 - \delta) + u^{(1)}(x), \qquad \phi^{(2)}(x) = x + u^{(2)}(x)
\end{equation}
The intra-chain energy density can be written as $\E_{\rm intra} \sim (\partial_x u^{(1)}(x))^2 + (\partial_x u^{(2)}(x))^2 \sim (\partial_x u^{\rm cm}(x))^2 + (\partial_x u^{(21)}(x))^2$ where $u^{\rm cm} = u^{(1)} + u^{(2)}$ and $u^{(21)} = u^{(2)} - u^{(1)}$. Furthermore, since $\phi(x)$ and $u(x)$ differ only by a linear function of $x$, we can write $\E_{\rm intra} \sim (\partial_x \phi^{\rm cm}(x))^2 + (\partial_x \phi^{(21)}(x))^2$ (up to a total derivative). The inter-chain energy is only a function of $\phi^{(21)} = \phi^{(2)} - \phi^{(1)}$. Thus, the relative and center of mass coordinates decouple and we can focus on the energy functional of the relative coordinate $\E[\phi^{(21)}(x)]$. For  notational simplicity, we will drop the superscript $(21)$ in what follows.

We can see now immediately that the energy is still invariant under the mismatch symmetry which acts through the shift $\phi(x) \mapsto \phi(x + \xi)$ since the energy does not depend on the matching point $O$ of the beating pattern encoded in $\phi(x)$. The action of the mismatch symmetry is however not simply a relative shift of one layer relative to the other. For example, for infinitesimal $\xi$, the symmetry action is $\phi(x) \mapsto \phi(x) + \xi \partial_x \phi(x)$. For the non-relaxed structure, $\phi(x) = x \delta$ and the symmetry simply acts as a shift $\phi(x) \mapsto \phi(x) + \xi \delta$. On the other hand, for the relaxed structure, the symmetry acts as $\phi(x) \mapsto \phi(x) + \xi (\delta + \partial_x u(x))$ which corresponds to a spatially dependent shift. 
It is important to emphasize that the symmetry remains a continuous local symmetry in the presence of relaxation.

Note that it is possible to also to define the symmetry in the lattice case away from the continuum limit provided that we retain the assumption that the relative and center of mass coordinates are decoupled. This is equivalent to the assumption that lattice relaxation only takes place in one of the chains, let's say the top one, with the other remaining rigid, i.e. $u_n^{(1)} = 0$. This model is called FK model \cite{ChaikinLubensky}. The do-nothing transformation which corresponds to a relabelling of the atoms in the top layers is now implemented by applying a spatially non-uniform shift
\begin{equation}
    \phi_n^{(2)} \mapsto \phi_{n+1}^{(2)} = \phi_n^{(2)} + b + u_{n+1}^{(2)} - u_{n}^{(2)}
    \label{phinShift}
\end{equation}
The mismatch symmetry corresponds to the shift $\phi_n^{(2)} \mapsto \phi_n^{(2)} + a$. Thus, the mismatch symmetry group is generated by the transformations
\begin{equation}
    \phi_n^{(2)} \mapsto \phi_n^{(2)} + N_1 a + N_2 b + u_{n+N_2}^{(2)} - u_{n}^{(2)}
    \label{phinSymm}
\end{equation}
modulo the transformations in (\ref{phinShift}). The symmetry transformation (\ref{phinSymm}) acts by shifting the matching point $O$ by $N_1$. The relaxed structure is commensurate if there exists $N_1$ and $N_2$ such that $N_1 a + u_{n+N_2}^{(2)} - u_{n}^{(2)} \mod b = 0$ for all $n$ and incommensurate otherwise. We can now apply the same arguments we used for the non-relaxed system showing that this symmetry is discrete for a commensurate system and arbitrarily close to a non-local continuous symmetry for the incommensurate case. The latter will again give rise to phason modes.

\subsubsection{Gapless excitations}\label{gapless_excitations}

By identifying the symmetries in the two-chain model, we can count the possible Goldstone modes corresponding to spontaneously breaking the continuous ones. First, there is the overall translation symmetry which simultaneously shifts both chains giving rise to the acoustic phonons. In addition, the lattice model has a phason mode only for $a/b \notin \mathbb{Q}$ which arises from breaking the continuous but non-local phase symmetry. As we explained earlier, this symmetry arises from the invariance of the energy under relative shifts of the two layers. On the other hand, the continuum model has a true Goldstone mode, ``the moir\'e phonon" arising from breaking the continuous local symmetry corresponding to shifting the origin of the moir\'e pattern by performing a (generally non-uniform) relative shift between the layers. Thus, the phason of the incommensurate lattice model becomes the moir\'e phonon of the lattice model. This means that all phasonic features of this mode should scale as $O(\delta) = O(a/L_M)$ relative to the moir\'e length/energy scale (see Appendix~\ref{correction_in_free_energy} for a detailed discussion). The key to validating this argument is a separation in length scales ($a/L_M \ll 1$), based on which the continuum limit is well-defined. In other words, the continuum limit allows us to treat a quasi-periodical lattice as a periodical one (the distinction between incommensurate and commensurate becomes vague), promoting a symmetry that is either discrete local or continuous non-local to be a continuous local symmetry giving rise to well-defined Goldstone modes.

\begin{figure}
    \centering
    \includegraphics[width=0.48\textwidth]{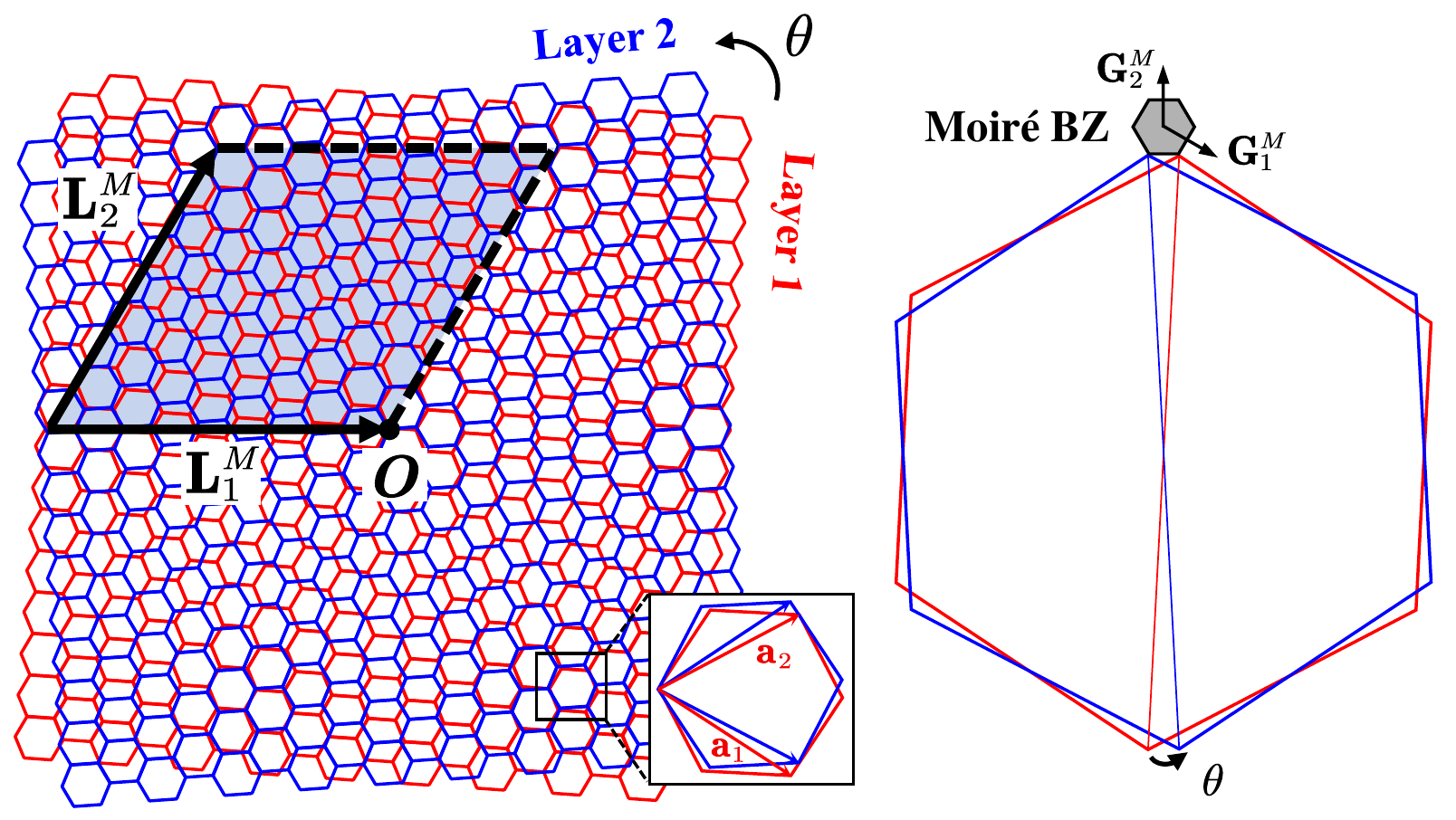}
    \caption{The schematic plots of the twisted bilayer graphene in real (left panel) and reciprocal (right panel) spaces.}
    \label{fig:TBG}
\end{figure}

\subsection{Correspondences between two-chain model and TBG}\label{SymmsInTBG}
\begin{table}[ht]
\begin{tabular}{|l|l|}
\hline
Two-Chain Model & Twisted Bilayer Graphene \\ \hline
\makecell[c]{Matching point $O$ }               & \makecell[c]{Twisting center $\boldsymbol{O}$ }                        \\\hline
\makecell[c]{$L_M = \frac{a}{1-a/b}$ }            &  \makecell[c]{$L_M = \frac{a}{2\sin(\theta/2)} $}                       \\\hline
\makecell[c]{$G_M = \left(1-\frac{a}{b}\right)\frac{2\pi}{a} $}              & \makecell[c]{$\bG^M_i = (1-\mathcal{R}_{\theta})\ba^*_i  $ }                     \\\hline
 \makecell[c]{$u^{(l)}(x)$  }             & \makecell[c]{$\bu^{(l)}(\br) $   }                    \\ \hline
\end{tabular}\caption{The correspondences between the two-chain model and the twisted bilayer graphene.}\label{table_correspondences}
\end{table}

The exact same story can be applied to the case of TBG as one can make the correspondences listed in Table~\ref{table_correspondences},
where for TBG, $a$ is the graphene lattice constant, $L_M$ is the moir\'e length scale, $\bG^M_i$ is the moir\'e reciprocal lattice vector, $ \mathcal{R}_{\theta}$ is the rotation matrix with a rotation angle $\theta$, and $\ba^*_i$ is the graphene reciprocal lattice vector. The geometry of the TBG is plotted in Fig.~\ref{fig:TBG}, where the real space and reciprocal space configurations are shown on the right and left panels, respectively. 

\begin{figure}
    \centering
    \includegraphics[width=0.45\textwidth]{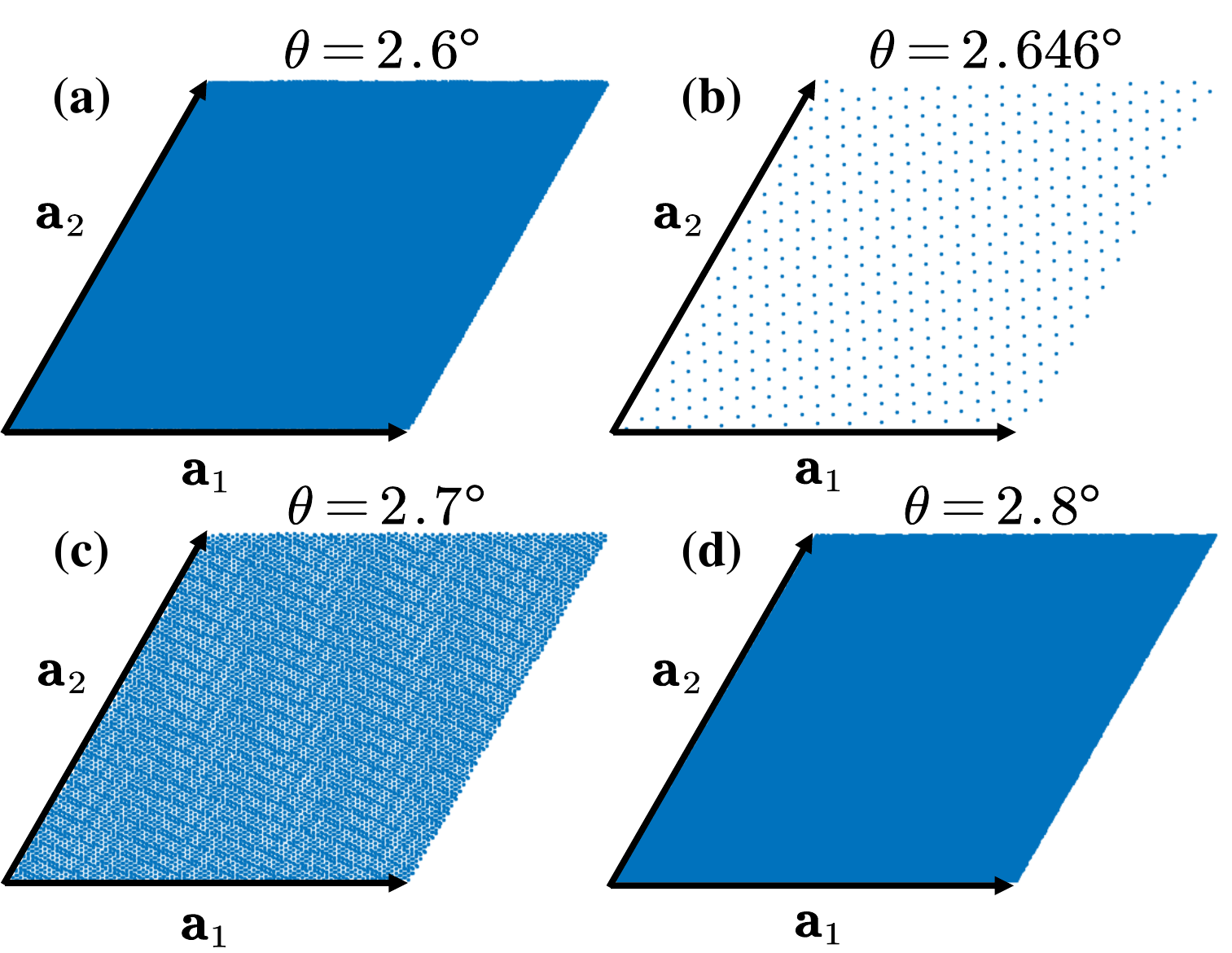}
    \caption{Folding $10^6$ graphene lattice points into the graphene unit cell ($\ba_1$ and $\ba_2$ are the two graphene lattice vectors) after a rotation for various angles: (a) $\theta=2.6^\circ$, (b) $\theta=2.646^\circ$, (c) $\theta=2.7^\circ$, and (d) $\theta=2.8^\circ$. Here $\theta= 2.646^\circ$ corresponds to a commensurate stacking of two graphene layers while the others do not.}
    \label{fig:PhSymm}
\end{figure}

The symmetry arguments for the two-chain model can be directly applied to the bilayer graphene both in the lattice model or continuum limit. For definiteness, we take one layer, let's say the top one, to be fixed and the other (bottom layer) to be rotated by an angle $\theta$. The mismatch symmetry group is given by translations of the top layer by a bottom-layer lattice vector $\R_\theta (n \ba_1 + m \ba_2)$ modulo translations by a top-layer lattice vector: $\{\mathcal{R}_{\theta}(n_1\ba_1+n_2\ba_2) \mod (\ba_1,\ba_2) | n_1,n_2\in \mathbb{Z} \} $. For commensurate stacking, this symmetry group is finite while for incommensurate stacking, this gives rise to a dense covering of the graphene uni cell. This is illustrated in Fig.~\ref{fig:PhSymm} showing $10^6$ different graphene lattice points ($1\leq n_1,n_2 \leq 10^3$) folded into the graphene unit cell for different twisting angles. We can see that the exact commensurate stacking leads to a discrete cover of the graphene unit cell while the incommensurate stacking leads to a dense cover.

In the continuum limit, similar to the two-chain model, the mismatch symmetry becomes a continuous local symmetry given by
\begin{equation}\label{TBGMisMatchSymm}
    \bu^{(21)}(\br)\mapsto\bu^{(21)}(\br-\bxi) + (\mathcal{R}^{-1}_{\theta}-1)\bxi,
\end{equation}
which is continuous since $\bxi\to \boldsymbol{0}$ can vary continuously. Here, as before, $\bu^{(21)}(\br)\equiv \bu^{(2)}(\br)-\bu^{(1)}(\br) $ is the relative displacement. Similar to the two chain model, one can use the new variable $\boldsymbol\phi(\br) = \bu^{(21)}(\br) + \boldsymbol{\eta}_0(\br)$ where  $\boldsymbol{\eta}_0 \equiv (1-\mathcal{R}_{\theta}^{-1})\br$ is the non-relaxed displacement, and then the symmetry action becomes simply a shift in its coordinate: $\boldsymbol\phi(\br)\mapsto\boldsymbol\phi(\br+\bxi)$ \cite{ochoa2019moire, ochoa2022degradation}.
The validity of taking the continuum limit is the separation of the length scales, i.e. $a/L_M \ll 1$, which corresponds to a small mismatch for the two-chain model ($1-a/b \ll 1$) or a small $\theta$ for TBG [$2\sin(\theta/2) \ll 1$].
\begin{figure}
    \centering
    \includegraphics[width=0.48\textwidth]{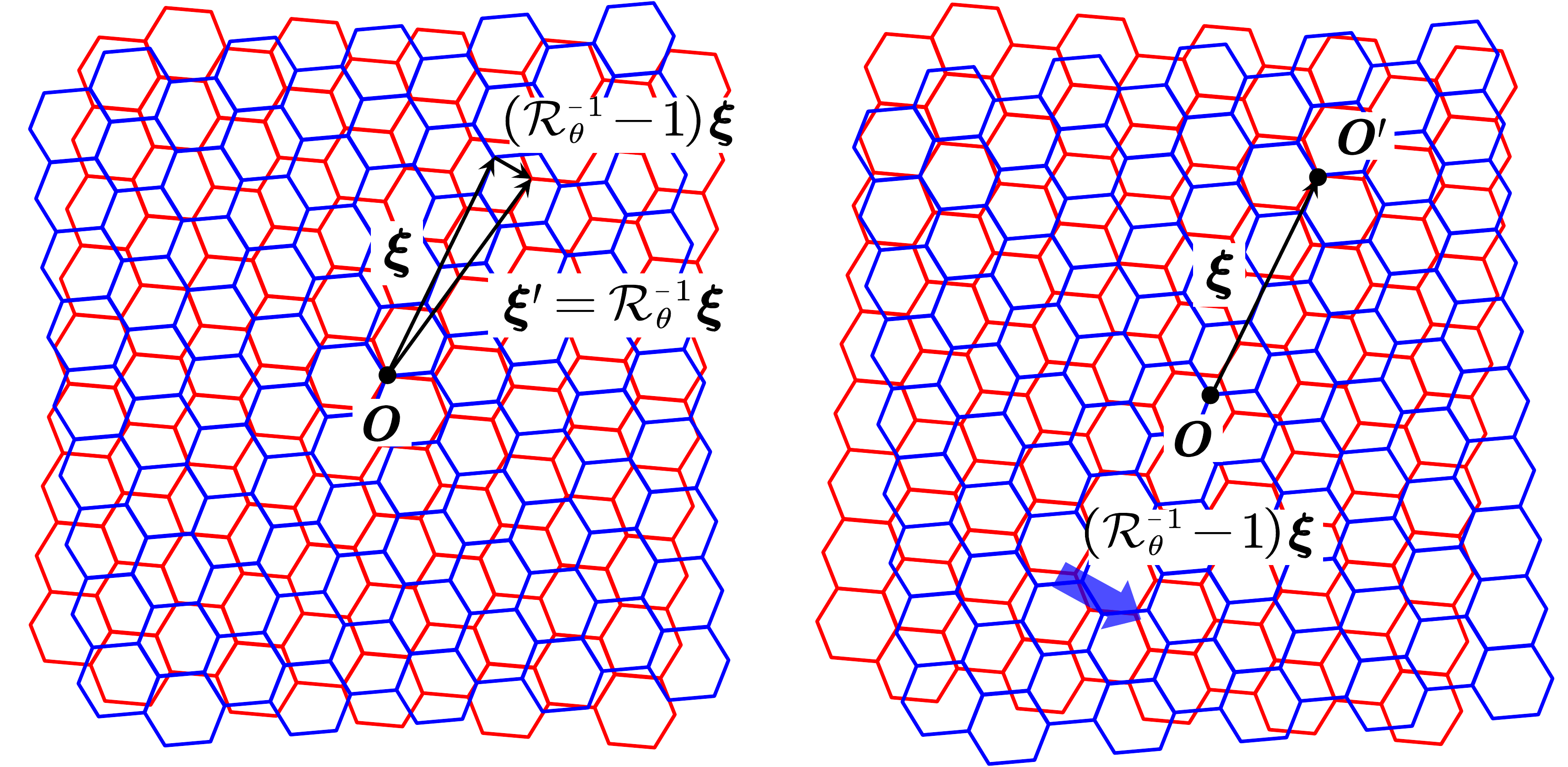}
    \caption{The illustration of the mismatch symmetry in TBG. (Left panel) the original stacking configuration with the twisting center $\bO$ and two lattice vectors $\bxi$ and $\bxi'$. (Right panel) the new configuration by shifting layer 2 by the difference of the two lattice vectors: $(\mathcal{R}_{\theta}^{-1}-1)\bxi$. As a result, the twisting center is shifted from point $\bO$ to $\bO'$.}
    \label{fig:MMSymm}
\end{figure}

We can see the symmetry explicitly by recalling the Lagrangian for the lattice dynamics in the continuum limit given in Ref.~\cite{nam2017lattice} that 
\begin{equation}\label{TBGLagrangian}
    \begin{split}
        &\mathcal{L}_{\text{TBG}}[\bu^{(1)},\bu^{(2)}] \\
        &=\sum_{l=1}^2 \left[\frac{\rho}{2}(\dot{\bu}^{(l)})^2 -  \frac{\lambda}{2}(\boldsymbol{\nabla}\cdot\bu^{(l)})^2 - \frac{\mu}{4}(\partial_i u_j^{(l)}+\partial_j u_i^{(l)})^2\right] \\
        &\quad- V^{(21)}[\br,\bu^{(21)}],
    \end{split}
\end{equation}
where $ \rho = 7.61\times 10^{-7}$ kg/m$^2$ is the area density of single-layer graphene, $\lambda\approx 3.25 $ eV/$\text{Å}^{2}$ and $\mu\approx 9.57$ eV/$\text{Å}^{2}$ are graphene’s Lam\'e factors~\cite{koshino2019moire}, and $V^{(21)}[\br, \bu^{(21)}]$ is the binding energy between two layers.

In general $V^{(21)}[\br, \bu^{(21)}]$ is only a function of the combination ${\boldsymbol \phi}(\br) = \bu^{(21)}(\br) + (1 - \R_\theta^{-1}) \br$ which is invariant under ${\boldsymbol \phi}(\br) \mapsto {\boldsymbol \phi}(\br) + \ba_i$, $i=1,2$ which means it can be expanded in terms of $\cos [(n \bb_1 + m \bb_2) \cdot {\boldsymbol \phi}(\br)]$ where $\bb_{1,2}$ are the graphene reciprocal lattice vector. The simplest choice of $V^{(21)}$ is to use the leading harmonics \cite{koshino2019moire}
\begin{equation}\label{TBGbindingEnenrgy}
    V^{(21)}[\br,\bu^{(21)}] = V^{(21)}[{\boldsymbol \phi}(\br)] = 2V_0\sum_{i=1}^3 \cos\left\{\bb^{(1)}_i \cdot {\boldsymbol \phi} (\br)\right\}
\end{equation}
Here, $\bb^{(1)}_1$ and $\bb^{(1)}_2$ are the reciprocal lattice vectors of the layer 1 and $\bb^{(1)}_3 = - \bb^{(1)}_1 -\bb^{(1)}_2$, and $V_0$ is the energy difference between ``AA" stacking and ``AB'' stacking graphene, where the latter is preferred energetically. $V_0$ is taken to be $161.5$ meV/nm$^2$~\footnote{The value for $V_0$ in this work is adopted from Ref.~\cite{koshino2019moire}. But we also notice a very different $V_0$ used in Ref.~\cite{ochoa2019moire}. However, using a different $V_0$ will not change the qualitative arguments in this work, instead, it will shift the numerics as it effectively rescales the twisting angle $\theta$.}. As one can see, the action in Eq.~\eqref{TBGMisMatchSymm} is indeed a symmetry of the Lagrangian above.

We notice that up a total derivative, the Lagrangian density can be written purely in terms of the center of mass coordinate $\bu^{\rm cm} = \bu^{(1)} + \bu^{(2)}$ and the phase variable ${\boldsymbol \phi}(\br)$:
\begin{equation}
\begin{split}
    \mathcal{L}_{\text{TBG}} = \mathcal{L}_{\rm cm}[\partial\bu^{\text{cm}}] + \mathcal{L}_{{\boldsymbol \phi}}[{\boldsymbol \phi}, \partial {\boldsymbol \phi}],
\end{split}
\end{equation}
The center of mass Lagrangian $\mathcal{L}_{\rm cm}[\partial\bu^{\text{cm}}]$ depends on on the gradients $\partial \bu^{\rm cm}$ and describes the graphene phonons:
\begin{equation}
\begin{split}
    &\mathcal{L}[\partial\bu^{\text{cm}}]  \\
    &= \frac{1}{2}\left[\frac{\rho}{2}(\dot{\bu}^{\text{cm}})^2 -  \frac{\lambda}{2}(\boldsymbol{\nabla}\cdot\bu^{\text{cm}})^2 - \frac{\mu}{4}(\partial_i u_j^{\text{cm}}+\partial_j u_i^{\text{cm}})^2\right],
\end{split}
\label{LTBGcmphi}
\end{equation}
whereas the relative phase Lagrangian is given by
\begin{equation}
\begin{split}
    &\mathcal{L}_{{\boldsymbol \phi}}[{\boldsymbol \phi}, \partial {\boldsymbol \phi}] = \frac{1}{2}\left[\frac{\rho}{2}(\dot{{\boldsymbol \phi}})^2 -  \frac{\lambda}{2}(\boldsymbol{\nabla}\cdot{\boldsymbol \phi})^2\right. \\
    &\qquad\left.- \frac{\mu}{4}(\partial_i \phi_j +\partial_j \phi_i)^2\right]- V^{(21)}[{\boldsymbol \phi}(\br)].
\end{split}
\end{equation}
We see that the Lagrangian does not depend explicitly on $\br$ which implies its invariance under translation $\br \mapsto \br + \boldsymbol \xi$. This is the continuum limit of the mismatch symmetry responsible for the moir\'e phonon. On the other hand, the fact that the Lagrangian depends only on the derivatives of the field $\bu^{\rm cm}$ implies invariance under $\bu^{\rm cm} \mapsto \bu^{\rm cm} + \boldsymbol \epsilon$ which gives rise to the graphene phonon. Thus, the system has 4 gapless Goldstone modes arising from the spontaneous breaking of 2 continuous symmetries in 2 dimensions.

\subsection{Generalization to TMG}\label{generalization}
\begin{figure}
    \centering
    \includegraphics[width=0.48\textwidth]{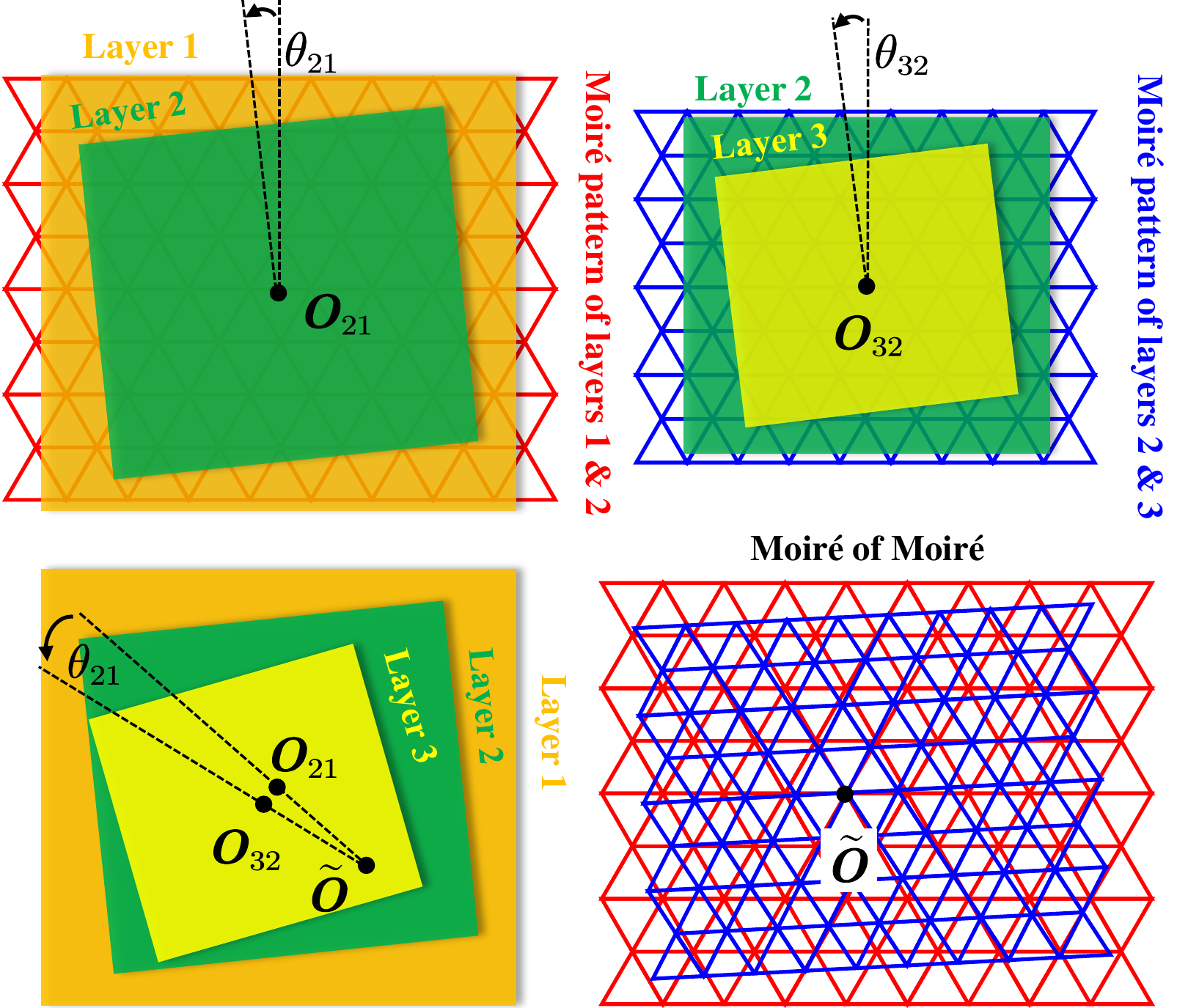}
    \caption{The schematic plots of the twisted trilayer graphene with consecutive twisted angles $\theta_{21}$ and $\theta_{32}$. The top left and top right panels show two different moir\'e patterns formed between layers 1 and 2 with a twisting center $\bO_{21}$ and between layers 2 and 3 with a twisting center $\bO_{32}$. The two corresponding triangular moir\'e superlattices are indicated by red and blue grids. In the bottom left panel, a generally twisted trilayer graphene is formed by twisting the top right moir\'e by an angle $\theta_{21}$ and then attaching it to the top left moir\'e with an exact match of layer 2. The bottom right panel shows the resulting beating pattern as a moir\'e of moir\'e with a secondary twisting center $\tilde \bO$.}
    \label{fig:TTG}
\end{figure}
The extension from TBG to TMG is quite natural since the symmetry arguments are also applicable to the system with more layers. The Lagrangian for the TMG is a direct generalization of that for the TBG:
\begin{equation}\label{TMGLagrangian}
    \begin{split}
        &\mathcal{L}_{\text{TMG}}[\bu,\partial\bu] =\sum_{l=1}^N \left[\frac{\rho}{2}(\dot{\bu}^{(l)})^2 -  \frac{\lambda}{2}(\boldsymbol{\nabla}\cdot\bu^{(l)})^2 \right.\\
        &\left.- \frac{\mu}{4}(\partial_i u_j^{(l)}+\partial_j u_i^{(l)})^2\right] - \sum_{l=1}^{N-1}V^{(l+1,l)}[\br,\bu^{(l+1,l)}],
    \end{split}
\end{equation}
where the inter-layer binding energies are defined as
\begin{equation}
\begin{split}
    &V^{(l+1,l)}[\br,\bu^{(l+1,l)}] \\
    &= 2V_0\sum_{i=1}^3 \cos\left\{\bb^{(1)}_i \cdot\left[\boldsymbol{\eta}^{(l+1,l)}_0 +  \bu^{(l+1,l)}(\br)\right]\right\}
\end{split}
\end{equation}
with $\boldsymbol{\eta}^{(l+1,l)}_0\equiv (\mathcal{R}_{\theta_{l,1}}^{-1} - \mathcal{R}_{\theta_{l+1,1}}^{-1})\br $ being the non-relaxed displacement between the $l^{\text{th}}$ and $(l+1)^{\text{th}}$ layers. Here $\theta_{l,m} $ is defined as the relative twisting angle between the $l^{\text{th}}$ and $m^{\text{th}}$ layers and $\theta_{l,l}\equiv 0$. The generalized Lagrangian is written in a form where the first layer (or bottom layer) is the reference layer. In writing down the Lagrangian, we assume that there is no coupling between second adjacent layers, i.e. $l^{\text{th}} $ and $(l+2)^{\text{th}} $ layers, given that the Van der Waals interactions between layers are weak enough. One now can see that the two continuous symmetries for TBG are also present in TMG, and that the system is invariant under the following transformations:
\begin{equation}\label{TMGSymm}
    \begin{split}
        \bu^{(l)} &\mapsto \bu^{(l)}+\boldsymbol\delta \text{  for }l=1,\cdots,N; \\
        \bu^{(l+1,l)}(\br) &\mapsto  \bu'^{(l+1,l)}(\br)\text{  for }l=1,\cdots,N-1,  
    \end{split}
\end{equation}
with
\begin{equation}
    \bu'^{(l+1,l)}(\br)= \bu^{(l+1,l)}(\br + \bxi) 
        + (\mathcal{R}_{\theta_{l,1}}^{-1} - \mathcal{R}_{\theta_{l+1,1}}^{-1})\bxi,
        \label{MultiMismatch}
\end{equation}
which will give two sets of Goldstone modes in the continuum limit which is valid if all twist angles $\theta_l$ are sufficiently small.

However in the general case where the twist angles $\theta_l$ do not have a simple relation, we also expect phason modes arising from the fact the moir\'e patterns formed between successive layers are generally incommensurate with each other resulting in a higher-order moir\'e pattern or ``moir\'e of moir\'e" \cite{zhu2020modeling, zhu2020twisted}. To make this explicit, let us consider the trilayer case. Given a general twist trilayer graphene (TTG), we can think about the resulting beating pattern as consisting of two patterns: one formed between layers 1 and 2 with matching center $\bO_{21}$ and moir\'e lattice vectors $\bL^{(21)}_i$ and the second formed between layers 2 and 3 with matching center $\bO_{32}$ and moir\'e lattice vectors $\bL^{(32)}_i$. The two moir\'e patterns form a moir\'e of moir\'e with a new matching center $\tilde \bO$ as illustrated in Fig.~\ref{fig:TTG}. The global mismatch symmetry defined in Eq.~\eqref{MultiMismatch} corresponds to the simultaneous shift of the two matching points $\bO_{21}$ and $\bO_{32}$ thus acting as a standard translation symmetry for the two moir\'e patterns. We can also define a higher-order mismatch symmetry by shifting one of the matching centers, let's say $\bO_{21}$ by the moir\'e lattice vector of the other moir\'e pattern $\bL^{(32)}$. The resulting mismatch symmetry group is $\{n \bL^{(32)}_1 + m \bL^{(32)}_2 \mod (\bL^{(21)}_1, \bL^{(21)}_2)\}$. Similar to our previous considerations, this symmetry group will be finite if the moir\'e lattice formed by $\bL^{(21)}_i$ and $\bL^{(32)}_i$ are commensurate and will provide a dense covering of the moir\'e unit cell if the two lattices are incommensurate. In the latter case, we obtain phason modes associated with the moir\'e of moir\'e. We want to emphasize here that the matching points $\bO_{21} $, $\bO_{32}$, and $\tilde \bO$ are not independent of each other. As shown in Fig.~\ref{fig:TTG}(c), the secondary matching point $\tilde \bO$ is uniquely determined by the positions of $\bO_{21} $ and $\bO_{32}$.

We note here the special case of alternating twist angle $\theta_{21} = -\theta_{32}$ for which the two moir\'e patterns are identical \cite{khalaf2019magic, park2021tunable} which is the analog of the two-chain model with equal lattice constants for which the mismatch symmetry group is trivial. Thus, instead of gapless phason modes, we will expect some gapped modes. We would like to emphasize here that unless we assume a special relationship between the two twist angles $\theta_{21} \approx -\theta_{32}$, we will not have a separation of scales between the moir\'e scale and that of the moir\'e of moir\'e. Thus, we expect the phason modes of the general trilayer system to be true phasons with diffusive dynamics. Thus, for general twisted $n$-layer systems where all twist angles are small and not related in a simple way, we expect $2n$ gapless modes consisting of 2 graphene phonons, 2 moir\'e phonons, and $2 (n-2)$ phasons.

In the following sections, we will first discuss the lattice dynamics in TBG using the continuum model and formulate a low-energy effective field theory showing that the moir\'e acoustic phonons are indeed Goldstone modes coming from the spontaneous breaking of the continuous local mismatch symmetry and demonstrating the absence of diffusive terms in the continuum model. We then turn to the simplest alternating twisted graphene: alternating twisted trilayer graphene (ATTG), showing that it inherits the acoustic modes from its TBG counterpart with an additional set of gapped modes and illustrating the basic idea of the higher order moir\'e pattern.

\section{Twisted Bilayer Graphene and Its Goldstone Modes}
Our starting point is TBG where we want to gain intuitions about the Goldstone modes due to the spontaneous breaking of the mismatch symmetry, then we will discuss twisted multilayer graphene with more than two layers.
\subsection{Lattice dynamics in TBG and continuous symmetry}\label{SymmTBG}
The lattice configuration of TBG is shown in Fig.~\eqref{fig:TBG}. The Lagrangian for the lattice vibration in the TBG is given by Eq.~\eqref{TBGMisMatchSymm}, which can be decoupled into a center of mass $\bu^{\rm cm}$ and a relative coordinate $\bu^{(21)}$ as in Eq.~\ref{LTBGcmphi}. We will prefer in this section to use the relative coordinate variable $\bu^{(21)}$ rather than the phase variable $\boldsymbol \phi$ to make the microscopic form of the symmetry manifest. The Lagrangian for the relative coordinate has the form
\begin{equation}
\begin{split}
    &\mathcal{L}_{12}[\bu^{(21)},\partial\bu^{(21)}] = \frac{1}{2}\left[\frac{\rho}{2}(\dot{\bu}^{(21)})^2 -  \frac{\lambda}{2}(\boldsymbol{\nabla}\cdot\bu^{(21)})^2\right. \\
    &\qquad\left.- \frac{\mu}{4}(\partial_i u_j^{(21)}+\partial_j u_i^{(21)})^2\right]- V^{(21)}[\br,\bu^{(21)}].
\end{split}
\label{Lagu21}
\end{equation} 
The Lagrangian for $\bu^{(21)}$ gives the equation for equilibrium configuration by letting the time variation to be zero:
\begin{equation}\label{EquPosBi}
    (\lambda + \mu)\boldsymbol{\nabla}(\boldsymbol{\nabla}\cdot\bu^{(21)}_0) + \mu\boldsymbol{\nabla}^2\bu^{(21)}_0 = 2\frac{\partial V^{(21)}}{\partial \bu^{(21)}}\bigg|_{\bu^{(21)}_0(\boldsymbol{r})}.
\end{equation}
And then we can perturb this equilibrium allowing it to vary in space and time: $\bu^{(21)}(\br,t) = \bu^{(21)}_0(\br) + \delta\bu^{(21)}(\br,t)$, which gives the equation of motion for the vibration modes:
\begin{equation}\label{ELDyn}
\begin{split}
    \rho\delta \ddot\bu^{(21)} =&(\lambda + \mu)\boldsymbol{\nabla}(\boldsymbol{\nabla}\cdot\delta\bu^{(21)}) + \mu\boldsymbol{\nabla}^2\delta\bu^{(21)} \\
    &- 2\delta\bu^{(21)}\cdot\frac{\partial^2 V^{(21)}}{\partial \bu^{(21)}\partial \bu^{(21)}}\bigg|_{\bu^{(21)}_0(\boldsymbol{r})}.
\end{split}
\end{equation}
Numerical calculations for the vibrational spectrum were obtained in Ref.~\cite{koshino2019moire}. 

We now discuss the continuous symmetry in this relative displacement vibration and its corresponding Goldstone modes. Recall Eq.~\eqref{TBGMisMatchSymm}:
\begin{equation}
    \bu^{(21)}(\br) \mapsto  \bu'^{(21)}(\br) := \bu^{(21)}(\br + \bxi) + (1- \mathcal{R}^{-1}_{\theta})\bxi,
    \label{Symm}
\end{equation}
whose action is equivalent to the shift $\br \mapsto \br+\bxi$ since $ V^{(21)}[\br, \bu'^{(21)}] =  V[\boldsymbol{\phi}(\br + \bxi)] = V^{(21)} [\br + \bxi, \bu^{(21)}]$. One should immediately recognize that such continuous transformation is the TBG version of the mismatch symmetry. Taking $\bxi$ to be infinitesimal, the symmetry is generated by the infinitesimal transformation: $\bu^{(21)}(\br) \mapsto \bu^{(21)}(\br) + \delta \bu^{(21)}(\br)$ where we find
\begin{equation}
     \delta \bu^{(21)}(\br) = (\boldsymbol{\xi}\cdot\boldsymbol{\nabla}) (\bu^{(21)} + \boldsymbol{\eta}_0) = (\boldsymbol{\xi}\cdot\boldsymbol{\nabla} )\bu^{(21)} + (1- \mathcal{R}^{-1}_{\theta})\boldsymbol{\xi}.
    \label{InfSymm}
\end{equation}
By noting that
\begin{equation}
    \boldsymbol{\nabla} V^{(21)}[\br, \bu^{(21)}] = [\boldsymbol{\nabla}( \boldsymbol{\eta}_0 + \bu^{(21)})]\cdot \frac{\partial}{\partial \bu^{(21)}}  V[\br, \bu^{(21)}] ,
\end{equation}
which follows since the potential $V^{(21)}$ only depends on the combination $\boldsymbol{\phi}(\br) = \boldsymbol{\eta}_0(\br) + \bu^{(21)}(\br)$. It is straightforward to see that the symmetry transformation generated by $\delta \bu^{(21)}(\br)$ leaves the equation of motion in \eqref{ELDyn} [and hence also the Lagrangian in  Eq.~\eqref{Lagu21}] invariant. Thus, if $\bu^{(21)}$ is an equilibrium configuration [a solution of Eq.~\eqref{EquPosBi}], then $\delta \bu^{(21)}$ in Eq.~\eqref{InfSymm} yields a zero mode. Note that the Lagrangian has translation symmetry but not full rotation symmetry (only a $D_6$ subgroup). This situation is to be contrasted with what happens in a regular crystal where the lattice forms out of a phase, e.g. a liquid, which is translationally and rotationally symmetric. This has important implications for the form of the effective field theory of the soft modes \cite{ochoa2019moire}.

Note that the symmetry Eq.~\eqref{Symm} vanishes for the equilibrium configuration $\bu_0^{(21)}(\br)$ at $\theta = 0$ which corresponds to AA or AB stacking. In this case, the equilibrium configuration is a constant $\bu_0^{(21)}(\br) = \textbf{Const.}$ and the infinitesimal shift $\delta \bu^{(21)}_0(\br)$ identically vanishes and does not yield a zero mode. This is compatible with what is known about the absence of such mode in AA or AB stacked systems. Physically, the model corresponds to shifting the pattern of domain walls formed by $\bu^{(21)}_0(\br)$ which only yields a non-trivial model if this pattern is not a constant.
This is identical to the discussion of Sec.~\ref{symmInTCM} in the two-chain model where the mismatch symmetry vanishes for $a=b$.

\subsection{Low energy effective field theory: massless Goldstone Bosons}\label{Goldstone_Boson}
After identifying the symmetry, we expect gapless Bosonic modes upon spontaneously breaking such symmetry through the equilibrium configuration which is generally not invariant under continuous translations. To derive an effective theory for the soft modes, we need to perform a change of variables from $\delta \bu^{(21)}(\br,t)$ to another variable $\bxi(\br,t)$ such that the soft modes correspond to slowly varying $\bxi$. The definition of the variable $\bxi$ can thus be motivated by the symmetry action (\ref{InfSymm}) as
\begin{multline}
     \delta \bu^{(21)}(\br, t) = (\boldsymbol{\xi}(\br, t)\cdot\boldsymbol{\nabla}) (\bu_0^{(21)}(\br) + \boldsymbol{\eta}_0(\br)) \\ = (\boldsymbol{\xi}(\br,t)\cdot\boldsymbol{\nabla} )\bu_0^{(21)}(\br) + (1- \mathcal{R}^{-1}_{\theta})\boldsymbol{\xi}(\br,t).
    \label{XiDef}
\end{multline}
Note that in contrast to Eq.~\eqref{InfSymm} where $\bxi$ is a constant, $\bxi$ here is a space-time dependent field. For constant $\bxi$, (\ref{XiDef}) reduces to the symmetry action (\ref{InfSymm}) which ensures that vanishing energy for this configuration. The Lagrangian can be written explicitly in terms of $\bxi$ to quadratic order as
\begin{equation}\label{newLag}
    \mathcal{L}[ \bxi,\partial\bxi] = \frac{1}{2}\mathcal{M}_{ij}\dot\xi_i\dot\xi_j - \frac{1}{2}(\mathcal{A}_{ijlm}\partial_i\xi_j\partial_l\xi_m + \mathcal{B}_{ijm}^{-}\xi_m\partial_i\xi_j)
\end{equation}
with coefficients $ \mathcal{M}_{ij}$, $ \mathcal{A}_{ijlm}$, and $\mathcal{B}_{ijm}^{-}$ defined in Appendix~\ref{variable_transformations}. It is worth noting that $\mathcal{B}_{ijm}^{-}$ is a total derivative of displacement $\bu$ and satisfies $\partial_i\mathcal{B}_{ijm}^{-}=0 $, which ensures the linear dispersion of the soft modes. The Euler-Lagrangian equation in terms of $\bxi$ are:
\begin{equation}\label{transferedEoM}
\begin{split}
    \mathcal{M}_{qj}\ddot\xi_j = \left( \mathcal{B}_{iqj}^{-} + \partial_l\mathcal{A}_{ijlq} \right)\partial_i\xi_j + \mathcal{A}_{ijlq} \partial_l\partial_i\xi_j,
\end{split}
\end{equation}
It now becomes more evident that $\bxi = \textbf{const.}$ is a zero mode of the system, which allows us to have gapless excitation upon perturbing the zero mode. 
We want to emphasize here
that the transformation from Eq.~\eqref{lagrangianDensity} to Eq.~\eqref{newLagrangian} is a point transformation which preserves the equation of motion (one can also do the transformation directly from Eq.~\eqref{ELDyn} to Eq.~\eqref{transferedEoM}). The advantage of such transformation is that it allows us to have an effective description of the lowest energy modes.

\begin{figure}
    \centering
    \includegraphics[width = 0.4
    \textwidth]{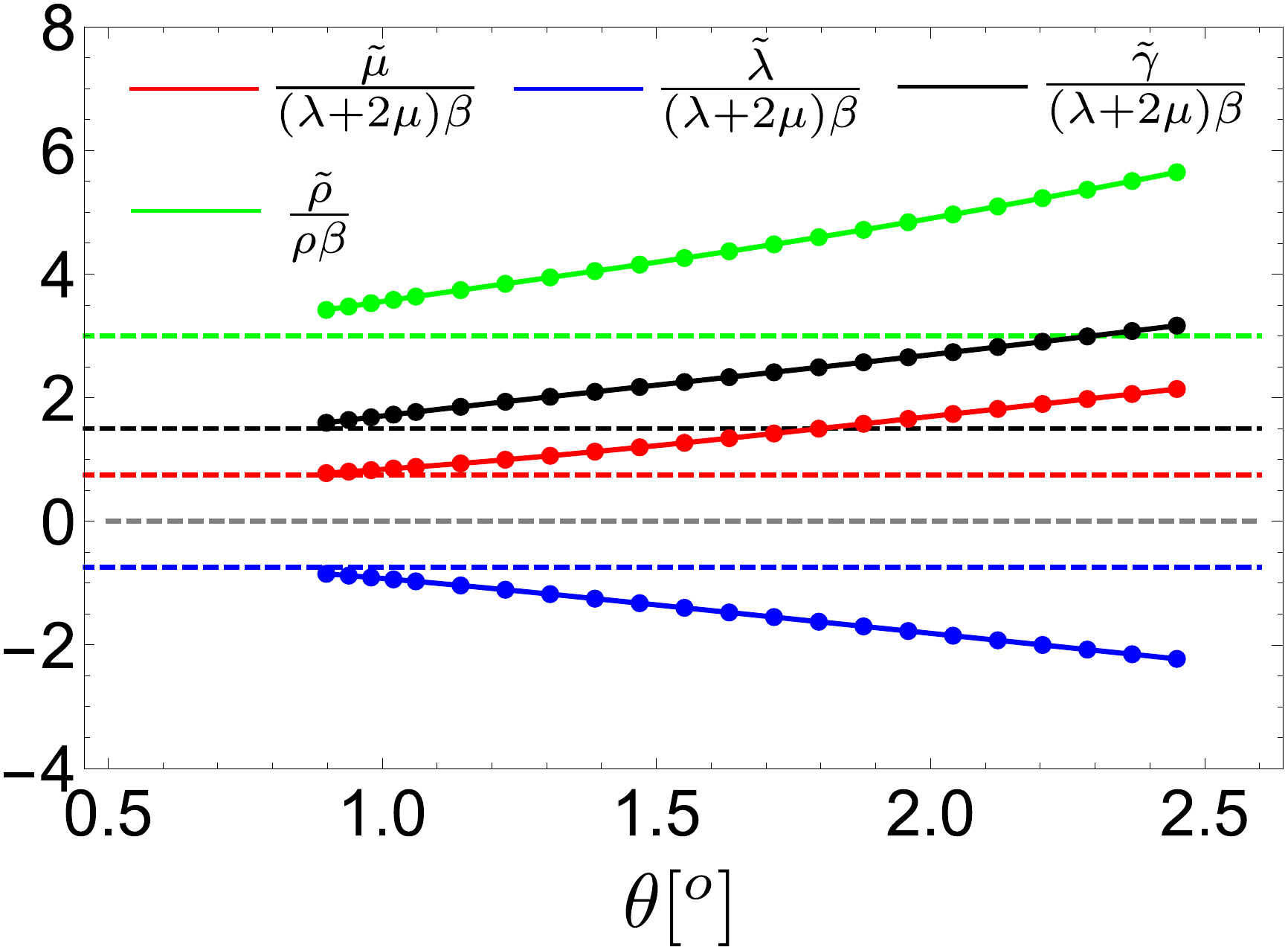}
    \caption{Parameters of the effective field theory: $\tilde \rho$ in terms of $\rho \beta$  and $\tilde \lambda$, $\tilde \mu$ and $\tilde \gamma$ in terms of $(\lambda + 2\mu) \beta$ where $\beta \propto \theta$ and is defined in Eq.~\ref{beta}. The small $\theta$ asymptotic behavior is indicated with dashed lines.}
    \label{fig:FTpars}
\end{figure}

Notice that the spatial dependencies of the coefficients $ \mathcal{M}_{ij}$, $ \mathcal{A}_{ijlm}$, and $\mathcal{B}_{ijm}^{-}$ will inevitably introduce coupling between slow varying and fast varying modes in $\bxi$. In order to obtain the effective field theory only for the low energy modes, we need to integrate out the fast varying modes in the action. A detailed treatment has been given in Appendix~\ref{EffFieldTheory}. The effective Lagrangian for the low energy modes reads
\begin{equation}
    \mathcal{L}_{\text{eff}} = \frac{1}{2}\Bar\M_{ij} \dot \xi_i  \dot \xi_j - \frac{1}{2} \tilde \A_{ijlm} \partial_i \xi_j \partial_l \xi_m,
\end{equation}
with
\begin{equation}
    \tilde \A_{ijlm} = \Bar\A_{ijlm} + \Delta_{ijlm},
\end{equation}
where $\Bar\M_{ij}$ and $ \Bar\A_{ijlm}$ are the spatial average of the coefficients $\M_{ij}$ and $\A_{ijlm}$, respectively, while $\Delta_{ijlm}$ defined in Eq.~\eqref{selfEnergy} contains information about the spatial variations of $ \mathcal{A}_{ijlm}$, and $\mathcal{B}_{ijm}^{-}$, giving the self-energy correction due to the coupling between slow varying and fast varying modes. We note here that a similar approach has been adopted in Ref.~\cite{ochoa2019moire} but without including the contribution from $\Delta_{ijlm}$. Due to $D_6$ symmetry, the effective Lagrangian can also be recast into the following form~\cite{ochoa2019moire}:
\begin{equation}
    \L_{\rm eff} = \frac{1}{2} \tilde \rho \dot {\boldsymbol \xi} \cdot \dot {\boldsymbol \xi} - \frac{1}{2} [\tilde \lambda (\nabla \cdot {\boldsymbol \xi})^2 + \frac{\tilde \mu}{2} (\partial_i \xi_j + \partial_j \xi_i)^2 + \tilde \gamma (\nabla \times {\boldsymbol \xi})^2 ]
\end{equation}
where the effective theory parameters are given by
\begin{gather}
    \tilde \rho = \frac{1}{2} \tr \M, \quad \tilde \lambda = \frac{1}{4} \tilde \A_{ijlm} (\sigma^0_{ij} \sigma^0_{lm} - \sigma^x_{ij} \sigma^x_{lm}), \nonumber \\ \tilde \mu = \frac{1}{4} \tilde \A_{ijlm} \sigma^x_{ij} \sigma^x_{lm},
    \qquad \tilde \gamma = - \frac{1}{4} \tilde \A_{ijlm} \sigma^y_{ij} \sigma^y_{lm},
\end{gather}
where $\sigma^\mu$, $\mu = 0,x,y,z$ are the Pauli matrices. Note the term $\tilde \gamma$ which explicitly breaks continuous rotation symmetry and would thus not appear in standard elasticity theory for crystals \cite{ochoa2019moire, FernandesSchmalian, ChaikinLubensky}. This leads to linearly dispersing modes with velocities \cite{ochoa2019moire}
\begin{equation}
    v_T = \sqrt{\frac{\tilde \gamma + \tilde \mu}{\tilde \rho}}, \qquad v_L = \sqrt{\frac{\tilde \lambda + 2 \tilde \mu}{\tilde \rho}}.
    \label{vTL}
\end{equation}

Let us now consider the limit of small $\theta$ where the equilibrium configuration $\bu_0^{(21)}(\br)$ takes the form of a sum of three domain wall solutions related by $2\pi/3$ rotations \cite{ochoa2019moire, koshino2019moire}. In terms of the variable $\bphi$, the domain wall parallel to the $y$-axis has the form has $\phi_x = $const. and $\phi_y$ depending only on $x$ and given explicitly by (see Appendix~\ref{smallTheta} for details)
\begin{equation}
    \phi^{\rm DW}_y(x) = \frac{\sqrt{3}}{\pi} \arctan \left(\frac{1}{\sqrt{3}} \tanh \sqrt{\frac{3}{2}} \frac{x}{2 w} \right),
\end{equation}
where $w$ is the length scale controlling the domain wall width given by
\begin{equation}
     w =  a\sqrt{\frac{3\sqrt{3}}{8\pi^2} \frac{\mu}{V_0}}.
\end{equation}
Using this domain wall solution, we can derive the leading asymptotics for the field theory parameters yielding
\begin{gather}
    \tilde \rho = 3 \beta \rho, \qquad \tilde \lambda = -\frac{3}{4} \beta (\lambda + 2 \mu), \nonumber \\ \tilde \mu = \frac{3}{4} \beta (\lambda + 2 \mu), \qquad \tilde \gamma = \frac{3}{2} \beta (\lambda + 2 \mu).
    \label{SmallthetaFTP}
\end{gather}
Here $\beta$ is given by
\begin{equation}
    \beta = \langle \phi_y'(x)^2 \rangle_{\rm MUC} = \frac{\sqrt{3} \theta a}{2 \pi^2 w} (\sqrt{6} - \frac{\sqrt{2}}{3}\pi)
    \label{beta}
\end{equation}
with $\langle f(\br) \rangle_{\rm MUC}$ denoting averaging the function $f(\br)$ over the moir\'e unit cell. We see that all the field theory parameters scale linearly with $\theta$ for small $\theta$ through the parameter $\beta$ so that the mode velocities in Eq.~\eqref{vTL} approach a constant. We also see that the three parameters $\tilde \lambda$, $\tilde \mu$, and $\tilde \gamma$ are all proportional to the same combination $\beta (\lambda + 2\mu)$ which implies the remarkable result that the ratio of the transverse and longitudinal velocities $v_T/v_L$ approaches the universal constant $\sqrt{3}$ which is independent on the material parameters as first noted in Ref.~\cite{koshino2019moire}. This  provides justification for the phenomenological spring lattice model proposed there. We note that our results differ from Ref.~\cite{ochoa2019moire} where
the ratio $v_T/v_L$ remains dependent on material parameters. The difference arises from the self-energy correction which we include and which was neglected in Ref.~\cite{ochoa2019moire}.

The field theory parameters $\tilde \rho$, $\tilde \lambda$, $\tilde \mu$, and $\tilde \gamma$ are plotted in Fig.~\ref{fig:FTpars} in appropriately scaled units. The results are consistent with the small $\theta$ asymptotics in Eq.~\eqref{SmallthetaFTP} and reproduce the numerical results of Ref.~\cite{koshino2019moire} for all angles, which implies that the moir\'e phonons are indeed the Goldstone modes from the spontaneous breaking of the mismatch symmetry in Eq.~\eqref{TBGMisMatchSymm}.

Since the gapless modes are true Goldstone modes arising from spontaneously breaking a continuous symmetry, they cannot acquire any mass or damping terms that remain finite for $\bq \rightarrow 0$. This is simply the statement that the amplitude of any scattering involving a Goldstone mode vanishes as the mode momentum $\bq$ vanishes \cite{WeinbergQFT}. This statement holds as long as we do not include terms beyond the continuum model. In particular, it implies that adding anharmonic interactions between the modes cannot lead to overdamping in contrast to the result of Ref.~\cite{ochoa2022degradation}. The reason for this discrepancy, as we discuss later, is that Ref.~\cite{ochoa2022degradation} computed the self-energy correction for the stacking variable $\delta {\boldsymbol \phi}$, which is not in general symmetry protected, rather than the Goldstone or phason variable $\bxi$. We can see this directly by including higher-harmonic terms in our model. Going beyond linear order, the definition of the $\bxi(\br)$ variable should be modified as \footnote{Note that this definition is not fixed uniquely by the requirement that $\bx_i = $const. is a zero mode. For instant, we could have instead chosen $\delta u_i^{(21)}({\bf r}) = (e^{\xi_j({\bf r}) \partial_j} - 1)u_{0i}^{(21)}({\bf r})$.} 
\begin{multline}
    \delta u_i^{(21)} = u_{0i}^{(21)}(\br + \bxi(\br)) - u_{0i}^{(21)}(\br) = \\
    \xi_j \partial_j (u_{0i}^{(21)} + \eta_{0i}) + \xi_j \xi_l \partial_j \partial_l u_{0i}^{(21)} + \dots 
    \label{NonLinearTransformation}
\end{multline}
The vertex function can be expanded to any order $n$ in $\xi$ as
\begin{equation}
    L^{(n)} = \sum_{\bq_1, \dots, \bq_n} F^{(n)}_{l_1,\dots,l_n}(\bq_1, \dots, \bq_n) \xi_{l_1,\bq_1} \dots \xi_{l_n,\bq_n}
\end{equation}
where $F^{(n)}$ is symmetric under any permutation of the indices $l_i \leftrightarrow l_j$ and $\bq_i \mapsto \bq_j$. The symmetry acts by simple shifting on the variable $\xi$: $\xi_i \mapsto \xi_i + \epsilon_i$. Substituting $\xi_{l,\bq} \mapsto \xi_{l,\bq} + \epsilon_l \delta_{\bq,0}$ and considering the leading order in $\epsilon$ implies that $F^{(n)}_{l_1,\dots,l_n}(\bq_1, \dots, \bq_n)$ vanishes whenever any $\bq_l = 0$. This establishes the absence of self-energy correction in the $\bq \rightarrow 0$ limit provided $F^{(n)}_{l_1,\dots,l_n}(\bq_1, \dots, \bq_n)$ is a continuous function of all $\bq_i$ which we assume. The vanishing of the third order term is shown explicitly in Appendix~\ref{thirdOrder}.
Note that since the variable transformation in Eq.~\eqref{NonLinearTransformation} is non-linear, it can also generate a non-trivial Jacobian when expanded beyond leading order. However, since the measure is invariant under the symmetry action, we can apply the same argument we did for the Lagrangian.

\section{Twisted Trilayer Graphene with Alternating Twisting Angles}
\subsection{Equilibrium position and mirror symmetry}\label{energtics}
We now consider the case of alternating twist trilayer graphene (ATTG). The discussion here is very similar to that for TBG with a few distinctions that we are going to emphasize. The Lagrangian for ATTG is obtained from Eq.~\eqref{TMGLagrangian} by letting $N=3$ and taking $\theta^{(21)} = -\theta^{(32)} = \theta$, with the binding energies given by:
\begin{equation}\label{bindingEnenrgy}
\begin{split}
    &V^{(21)}[\boldsymbol{r},\bu^{(21)}(\boldsymbol{r})] \\
    &= 2V_0\sum_{i=1}^3 \cos\left\{\ba^{*(1)}_i \cdot\left[(1 - \mathcal{R}_{\theta}^{-1})\br +  \bu^{(21)}(\br)\right]\right\}
\end{split}
\end{equation}
and 
\begin{equation}
\begin{split}
    &V^{(32)}[\boldsymbol{r},\bu^{(32)}(\boldsymbol{r})] \\
    &= 2V_0\sum_{i=1}^3 \cos\left\{\ba^{*(1)}_i \cdot\left[( \mathcal{R}_{\theta}^{-1}-1)\br +  \bu^{(32)}(\br)\right]\right\}.
\end{split}
\end{equation}
Again, we are assuming no direct coupling between layer 1 and layer 3 as the inter-layer interaction is presumably small and short-ranged, and we adopt the configuration where the system becomes ``AAA" (``ABA" will be similar) stacking when $\theta=0$.
Thus the equilibrium position is obtained through the Euler-Lagrangian equation for the static part, which gives the following differential equations for the three layers:
\begin{equation}
    \begin{split}
        (\lambda + \mu)\boldsymbol{\nabla}(\boldsymbol{\nabla}\cdot\bu^{(1)}) + \mu\boldsymbol{\nabla}^2\bu^{(1)} &= -\frac{\partial V^{(21)}}{\partial \bu^{(21)}}, \\
        (\lambda + \mu)\boldsymbol{\nabla}(\boldsymbol{\nabla}\cdot\bu^{(2)}) + \mu\boldsymbol{\nabla}^2\bu^{(2)} &= \frac{\partial V^{(21)}}{\partial \bu^{(21)}}-\frac{\partial V^{(32)}}{\partial \bu^{(32)}}, \\
        (\lambda + \mu)\boldsymbol{\nabla}(\boldsymbol{\nabla}\cdot\bu^{(3)}) + \mu\boldsymbol{\nabla}^2\bu^{(3)} &= \frac{\partial V^{(32)}}{\partial \bu^{(32)}} .
    \end{split}
\end{equation}
Instead of using $\bu^{(l)}$ ($l=1,2,3$) as the dynamical variables, it is more convenient to use the relative displacements $\bu^{(21)}$, $\bu^{(32)}$, and the center of mass movement $\bu^{\text{cm}}\equiv \bu^{(1)}+\bu^{(2)}+\bu^{(3)}$. The differential equations then become
\begin{equation}\label{equilibriumTTG}
    \begin{split}
        (\lambda + \mu)\boldsymbol{\nabla}(\boldsymbol{\nabla}\cdot\bu^{\text{cm}}) + \mu\boldsymbol{\nabla}^2\bu^{\text{cm}} &=0, \\
        (\lambda + \mu)\boldsymbol{\nabla}(\boldsymbol{\nabla}\cdot\bu^{(21)}) + \mu\boldsymbol{\nabla}^2\bu^{(21)} &= 2\frac{\partial V^{(21)}}{\partial \bu^{(21)}}-\frac{\partial V^{(32)}}{\partial \bu^{(32)}}, \\
        (\lambda + \mu)\boldsymbol{\nabla}(\boldsymbol{\nabla}\cdot\bu^{(32)}) + \mu\boldsymbol{\nabla}^2\bu^{(32)} &= 2\frac{\partial V^{(32)}}{\partial \bu^{(32)}} - \frac{\partial V^{(21)}}{\partial \bu^{(21)}},
    \end{split}
\end{equation}
which appears more symmetric. The center of mass movement $\bu^{\text{cm}}$ follows the identical equations as the displacement of single-layer graphene and thus has the same dynamics which give rise to the normal acoustic phonon modes. Our focus will be the relative displacements, $\bu^{(21)}$ and $\bu^{(32)}$.

The alternating angle in ATTG imposes a symmetry constraint on the equilibrium configurations. In Appendix~\ref{symmATMG}, we discuss a rotation adapted mirror symmetry in the alternating twisted multilayer graphene. In the case with an odd number of layers, such symmetry becomes local. For example, in the ATTG, we should have $\bu^{(21)} = -\bu^{(32)}$ [or equivalently, $\bu^{(1)}(\boldsymbol{r})=\bu^{(3)}(\boldsymbol{r})$]. One can think about this symmetry as the mirror symmetry in the $z$-direction with respect to the middle layer, i.e. a symmetry under the exchange of layer 1 and layer 3. Therefore, the differential equations become simply:
\begin{equation}\label{EquPos}
    (\lambda + \mu)\boldsymbol{\nabla}(\boldsymbol{\nabla}\cdot\bu^{(21)}) + \mu\boldsymbol{\nabla}^2\bu^{(21)} = 3\frac{\partial V^{(21)}}{\partial \bu^{(21)}},
\end{equation}
by noting that $V^{(21)}[\boldsymbol{r},\bu^{(21)}(\boldsymbol{r})]=V^{(32)}[\boldsymbol{r},\bu^{(32)}(\boldsymbol{r})]$. We note here that the equality between those two binding energies relies on the fact that the energy itself is an even functional of its arguments, i.e. $ V[\br,\bu] = V[-\br,-\bu]$. This then requires the binding energy to be isotropic at least in the continuum limit.

\subsection{Lattice vibrations in ATTG}
\begin{figure}
    \centering
    \includegraphics[width=0.48\textwidth]{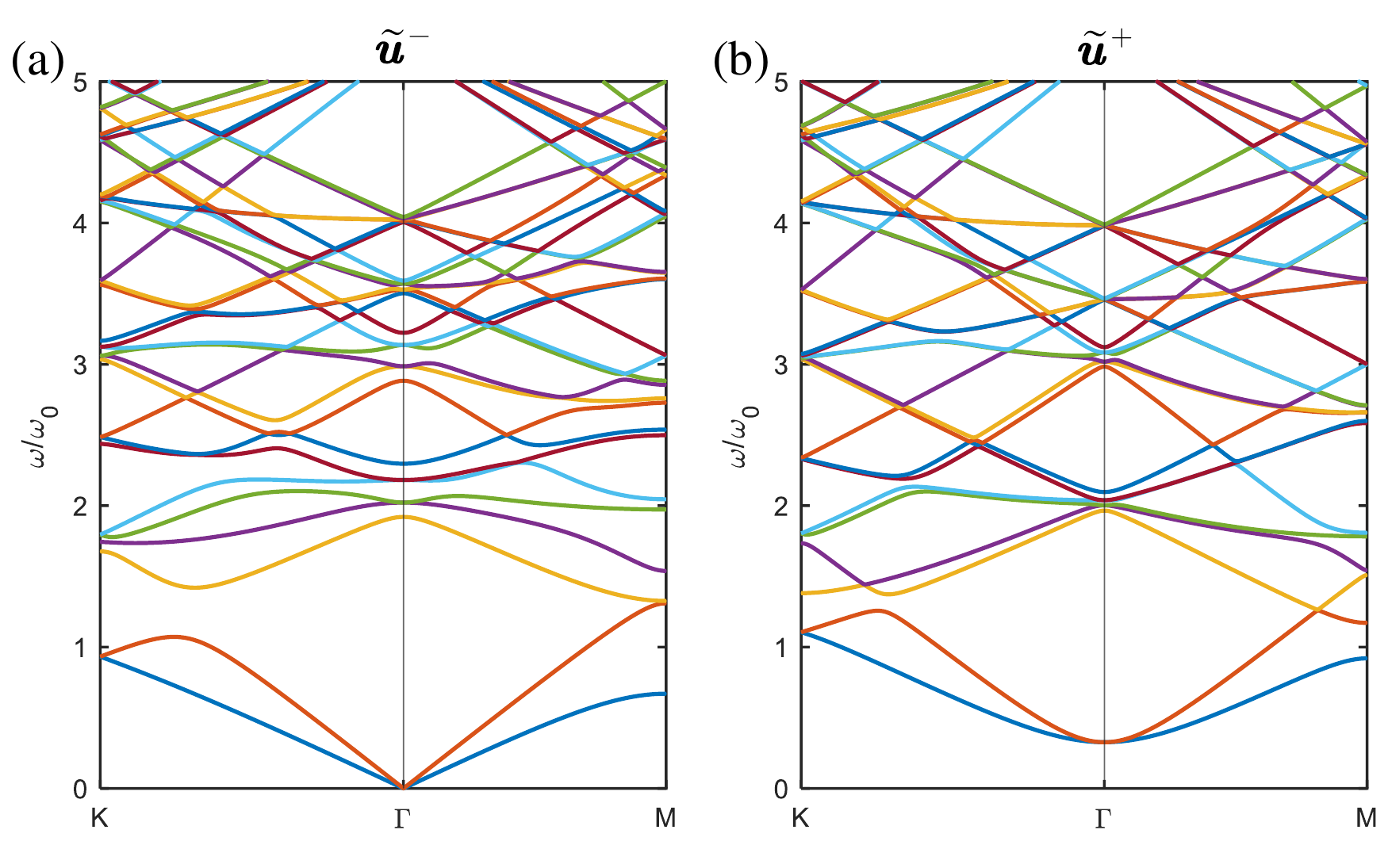}
    \caption{The vibrational spectrum of the alternating twisted trilayer graphene with twisting angle $\theta=2.65^{\circ}$ for (a) the mirror-symmetric variation $\tilde{\bu}^{-}$ and (b) the mirror anti-symmetric variation $\tilde{\bu}^{+}$. $\omega_0 = 2\pi\sqrt{\lambda/\rho}/L_M$ is the reference energy.}
    \label{fig:ATTG_spectrum}
\end{figure}

\begin{figure}
    \centering
    \includegraphics[width=0.45\textwidth]{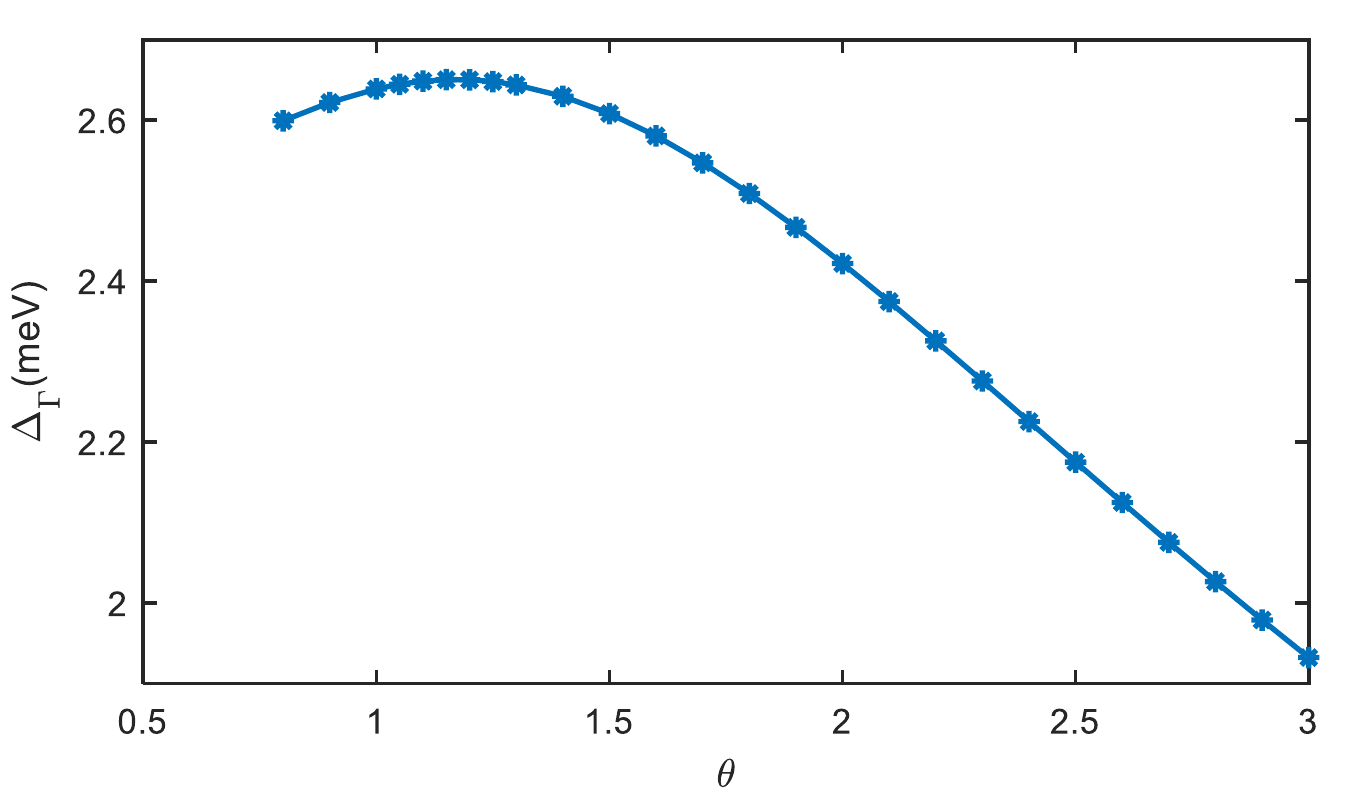}
    \caption{The gap $\Delta^{\text{ATTG}}_{\text{odd}}$ at the $\Gamma$ point for the vibrational mode $\tilde\bu^{+}$ as a function of the twisting angle $\theta$.}
    \label{fig:gap_ATTG}
\end{figure}

The equations of motion for the relative displacements $\bu^{(21)}$ and $\bu^{(32)}$ can be deduced from the Euler-Lagrangian equation:
\begin{equation}\label{EOM}
    \begin{split}
        \rho\ddot{\bu}^{(21)} =&(\lambda + \mu)\boldsymbol{\nabla}(\boldsymbol{\nabla}\cdot\bu^{(21)}) + \mu\boldsymbol{\nabla}^2\bu^{(21)} \\
        &- 2\frac{\partial V^{(21)}}{\partial \bu^{(21)}}+\frac{\partial V^{(32)}}{\partial \bu^{(32)}}, \\
        \rho\ddot{\bu}^{(32)} =&(\lambda + \mu)\boldsymbol{\nabla}(\boldsymbol{\nabla}\cdot\bu^{(32)}) + \mu\boldsymbol{\nabla}^2\bu^{(32)}\\
        &- 2\frac{\partial V^{(32)}}{\partial \bu^{(32)}} + \frac{\partial V^{(21)}}{\partial \bu^{(21)}}.
    \end{split}
\end{equation}
To see the vibration modes, we need to perturb the displacements, $\bu^{(21)}$ and $\bu^{(32)}$ near their equilibrium positions, namely letting $\bu^{(21)}(\boldsymbol{r},t) = \bu^{(21)}_0(\boldsymbol{r}) + \delta\bu^{(21)}(\boldsymbol{r},t)$ and $\bu^{(32)}(\boldsymbol{r},t)=\bu^{(32)}_0(\boldsymbol{r})+\delta\bu^{(32)}(\boldsymbol{r},t)$ with $\bu^{(21)}_0(\boldsymbol{r})$ and $\bu^{(32)}_0(\boldsymbol{r})$ being the corresponding equilibrium displacements. Expanding the last two equations in Eq.~\eqref{EOM}, we get
\begin{equation}\label{EOMnew}
    \begin{split}
        \rho\delta\ddot{\bu}^{(21)} =&(\lambda + \mu)\boldsymbol{\nabla}(\boldsymbol{\nabla}\cdot\delta\bu^{(21)}) + \mu\boldsymbol{\nabla}^2\delta\bu^{(21)}\\
        &- 2\delta\bu^{(21)}\cdot\frac{\partial^2 V^{(21)}}{\partial \bu^{(21)}\partial \bu^{(21)}}\bigg|_{\bu^{(21)}_0(\boldsymbol{r})}\\
        &+\delta\bu^{(32)}\cdot\frac{\partial^2 V^{(32)}}{\partial \bu^{(32)}\partial \bu^{(32)}}\bigg|_{\bu^{(32)}_0(\boldsymbol{r})}, \\
        \rho\delta\ddot{\bu}^{(32)} =& (\lambda + \mu)\boldsymbol{\nabla}(\boldsymbol{\nabla}\cdot\delta\bu^{(32)}) + \mu\boldsymbol{\nabla}^2\delta\bu^{(32)}\\
        &- 2\delta\bu^{(32)}\cdot\frac{\partial^2 V^{(32)}}{\partial \bu^{(32)}\partial \bu^{(32)}}\bigg|_{\bu^{(32)}_0(\boldsymbol{r})} \\
        &+ \delta\bu^{(21)}\cdot\frac{\partial^2 V^{(21)}}{\partial \bu^{(21)}\partial \bu^{(21)}}\bigg|_{\bu^{(21)}_0(\boldsymbol{r})}.
    \end{split}
\end{equation}
One should notice that by the same argument that $\bu^{(21)}= -\bu^{(32)}$, we have
\begin{equation}\label{constraint}
    \frac{\partial^2 V^{(21)}}{\partial \bu^{(21)}\partial \bu^{(21)}}\bigg|_{\bu^{(21)}_0(\boldsymbol{r})}=\frac{\partial^2 V^{(32)}}{\partial \bu^{(32)}\partial \bu^{(32)}}\bigg|_{\bu^{(32)}_0(\boldsymbol{r})}.
\end{equation}
Let's introduce the mirror even-odd displacements $\tilde{\bu}^{\pm} \equiv  \delta\bu^{(21)}\pm\delta \bu^{(32)}$. Then the equations of motion can be further decoupled as:
\begin{equation}\label{vibMode}
    \begin{split}
        \rho\ddot{\tilde{\bu}}^{+} =&(\lambda + \mu)\boldsymbol{\nabla}(\boldsymbol{\nabla}\cdot\tilde{\bu}^{+}) + \mu\boldsymbol{\nabla}^2\tilde{\bu}^{+}\\
        &- \tilde{\bu}^{+}\cdot\frac{\partial^2 V^{(21)}}{\partial \bu^{(21)}\partial \bu^{(21)}}\bigg|_{\bu^{(21)}_0(\boldsymbol{r})}, \\
        \rho\ddot{\tilde{\bu}}^{-} =&(\lambda + \mu)\boldsymbol{\nabla}(\boldsymbol{\nabla}\cdot\tilde{\bu}^{-}) + \mu\boldsymbol{\nabla}^2\tilde{\bu}^{-}\\
        &- 3\tilde{\bu}^{-}\cdot\frac{\partial^2 V^{(21)}}{\partial \bu^{(21)}\partial \bu^{(21)}}\bigg|_{\bu^{(21)}_0(\boldsymbol{r})}.
    \end{split}
\end{equation}
Under the mirror symmetry action, we see that the $\tilde{\bu}^{+}$ is odd while the $\tilde{\bu}^{-}$ is even. We plot the spectrum corresponding to those two modes separately in Fig.~\ref{fig:ATTG_spectrum} for a twisting angle $\theta=2.65^{\circ}$. As one can see, the mirror-symmetric variation provides two gapless modes, while the mirror anti-symmetric variation opens gaps at the $\Gamma$ point. We also plot in Fig.~\ref{fig:gap_ATTG} the gap at the $\Gamma$ point for the vibration mode $\tilde{\bu}^{+}$ as a function of the twisting angle, which shows clearly that the gap is peaked at $\theta\approx1.2^\circ$. Here we want to emphasize that this $1.2^\circ$ is obtained under a specific potential constant $V_0$ used in this work and we argue that it will shift to $\sqrt{\alpha}\times1.2^\circ$ if a new $\tilde V_0 = \alpha V_0$ is adopted.

In the next two sections, we will demonstrate that the two gapless modes are inherited from the TBG counterpart of the ATTG, which are the Goldstone modes with the same origin, while the gaped modes are consequences of breaking the mirror symmetry.

\subsection{Reduction to lattice dynamics in TBG}
The differential equation Eq.~\eqref{EquPos} has exactly the same structure as the corresponding TBG with the same twisted angle [see Eq.~\eqref{EquPosBi}].
One should notice that there is a subtle difference between the prefactors on the right-hand sides of Eq.~\eqref{EquPos} and Eq.~\eqref{EquPosBi}. We can replace the potential constant $V_0$ with a new constant $V'_0 = \frac{3}{2}V_0$ we have for the trilayer system, the equilibrium displacement follows:
\begin{equation}\label{EquPosNew}
    (\lambda + \mu)\boldsymbol{\nabla}(\boldsymbol{\nabla}\cdot\bu^{(21)}) + \mu\boldsymbol{\nabla}^2\bu^{(21)} = 2\frac{\partial V'^{(21)}}{\partial \bu^{(21)}},
\end{equation}
with
\begin{equation}
\begin{split}
    &V'^{(21)}[\boldsymbol{r},\bu^{(21)}(\boldsymbol{r})]\\
    &= 2V'_0\sum_{i=1}^3 \cos\left\{\ba^{*(1)}_i \cdot\left[(1 - \mathcal{R}_{\theta}^{-1})\br +  \bu^{(21)}(\br)\right]\right\}.
\end{split}
\end{equation}
In Sec.~\ref{Goldstone_Boson}, we discussed the domain wall formation in the TBG after the lattice relaxation. Here, we would also expect domain walls in the relaxed ATTG.
The domain wall width in the ATTG is given by
\begin{equation}
    w_d^{\text{ATTG}} \approx a\sqrt{\frac{3\sqrt{3}}{8\pi^2} \frac{\mu}{V'_0}} = \sqrt{\frac{2}{3}}w_d^{\text{TBG}}
\end{equation}
which is smaller than the width of the domain walls in a TBG for the same twist angle. In the continuum model, such rescaling factor can also absorb into the twisting angle $\theta$, based on which the mapping from ATTG to TBG is given by:
\begin{equation}
    \theta^{\text{ATTG}}\mapsto\theta^{\text{TBG}} = \sqrt{\frac{2}{3}}\theta^{\text{ATTG}},\quad \br \mapsto \br'=\sqrt{\frac{2}{3}}\br.
\end{equation}

Using the rescaled potential constant, Eq.~\eqref{vibMode} can also be rewritten as
\begin{equation}\label{EoMTri}
    \begin{split}
        \rho\ddot{\tilde{\bu}}^{+} =&(\lambda + \mu)\boldsymbol{\nabla}(\boldsymbol{\nabla}\cdot\tilde{\bu}^{+}) + \mu\boldsymbol{\nabla}^2\tilde{\bu}^{+} \\
        &- \frac{2}{3}\tilde{\bu}^{+}\cdot\frac{\partial^2 V'^{(21)}}{\partial \bu^{(21)}\partial \bu^{(21)}}\bigg|_{\bu^{(21)}_0(\boldsymbol{r})}, \\
        \rho\ddot{\tilde{\bu}}^{-} =&(\lambda + \mu)\boldsymbol{\nabla}(\boldsymbol{\nabla}\cdot\tilde{\bu}^{-}) + \mu\boldsymbol{\nabla}^2\tilde{\bu}^{-}\\
        &- 2\tilde{\bu}^{-}\cdot\frac{\partial^2 V'^{(21)}}{\partial \bu^{(21)}\partial \bu^{(21)}}\bigg|_{\bu^{(21)}_0(\boldsymbol{r})},
    \end{split}
\end{equation}
We see that the equation of motion for the mirror-even mode $\tilde{\bu}^{-}$, which corresponds to the simultaneous  movement of the top and the bottom layer relative to the middle layer, is identical to the relative displacement mode in TBG [see Eq.~\eqref{ELDyn}]. Thus $\tilde{\bu}^{-}$ has exactly the same spectrum as in TBG up to a re-scaling. On the other hand, the mirror-odd mode has no counterpart in TBG and corresponds to the movement of the top layer relative to the bottom layer with the middle layer fixed: $\bu^{(3)}(\boldsymbol{r})-\bu^{(1)}(\boldsymbol{r})$. Notice that in the equilibrium configuration we have $\bu_0^{(3)}(\boldsymbol{r})-\bu_0^{(1)}(\boldsymbol{r}) \equiv 0$ by the symmetry restriction. 

It is worth mentioning that this mapping from ATTG to TBG does not have a straightforward generalization to alternating twist multilayer systems with more than three layers due to the complicated form of the Euler-Lagrange equations in that case. This is to be contrasted with the mapping of the electronic spectrum which applies for alternating twist multilayer systems with an arbitrary number of layers \cite{khalaf2019magic, Ledwith2021tb}.

\subsection{Symmetry arguments in ATTG}
As discussed above, the ATTG hosts 4 gapless modes: 2 from the center of mass motion $\bu^{\text{cm}}$ and 2 from the mirror-symmetric displacement $\tilde{\bu}^{-}$. Unlike the generally twisted trilayer graphene where we expect 6 gapless modes in total, the two extra would-be Goldstone modes become gapped as shown in Fig.~\ref{fig:ATTG_spectrum}(b) where the lowest two bands for vibration modes $\tilde{\bu}^{+}$ have a gap at the $\Gamma$ point.

To explain the gapped modes, we can apply the same variable transformations as what we have done for TBG:
\begin{equation}
    \begin{split}
        \delta \bu^{(21)}(\br) & = (\boldsymbol{\xi}^{(21)}\cdot\boldsymbol{\nabla} )\bu^{(21)} + (1- \mathcal{R}^{-1}_{\theta})\boldsymbol{\xi}^{(21)}, \\
        \delta \bu^{(32)}(\br) & = (\boldsymbol{\xi}^{(32)}\cdot\boldsymbol{\nabla} )\bu^{(32)} + ( \mathcal{R}^{-1}_{\theta}-1)\boldsymbol{\xi}^{(32)},
    \end{split}
\end{equation}
and thus $\tilde{u}^{\pm}_{i} = \mathcal{C}_{ij}\xi_{j}^{\mp}$, where $\mathcal{C}_{ij}\equiv\partial_ju^{(21)}_i + (1- \mathcal{R}^{-1}_{\theta})_{ij} $ is the same generating tensor as in the TBG, and $\bxi^{\pm}\equiv\bxi^{(21)}\pm \bxi^{(32)}$. The continuous symmetry in Eq.~\eqref{TMGSymm} is then translated to $\bxi^{+}=\textbf{Const.}$ and $\bxi^{-} =\boldsymbol 0$, which means the mirror-symmetric part of the variation will give two gapless modes. However, the ground state configuration of the lattice deformation in the ATTG should preserve the mirror symmetry, i.e. the global minimum of the free energy lies exactly in the $\bxi^{-}=\boldsymbol 0$ sub-parameter space. Any variation away from the mirror-symmetric configuration costs energy and thus introduces masses to the vibration modes corresponding to $\tilde{\bu}^{+}$. 

Another way to think about the gaped modes is to take the ATTG as an ``AA'' stacking of two exactly same moir\'e patterns. Recall the discussion about the TMG in Sec.~\ref{generalization}, a twisted trilayer layer graphene can be viewed as the moir\'e of moir\'e patters. By analogy to the ``AA'' stacking bilayer graphene where one would get two acoustic phonon modes as in the single layer graphene and two additional gaped phonon modes, in ATTG, we get exactly two gapless modes inherited from the corresponding TBG but with re-scaled potential energy and two extra gaped modes by breaking the mirror symmetry. However, the moir\'e of moir\'e can also admit an ``AB" stacking of two same moir\'e patterns, which corresponds to an ATTG with ``AAB" stacking of three graphene layers as its $\theta =0$ limit. In such case, the mirror symmetry does not exist and neither does the reduction to the TBG, but the count for gapless modes is the same as the ``AA'' moir\'e of moir\'e patters.

\section{Discussions and Remarks}
In this work, we have identified the origin of the soft lattice vibration modes in twisted multilayer moir\'e heterostructures. Starting from a simple two-chain model, then generalizing to twisted bilayer and multilayer systems, we have identified a microscopic symmetry, which we dub the mismatch symmetry, which plays a pivotal role in determining the nature of these collective modes. For commensurate stacking, this symmetry is discrete and does not give rise to gapless modes whereas for incommensurate stacking, it is continuous but non-local and gives rise to phason modes with diffusive dynamics similar to those observed in quasi-periodic systems \cite{lubensky1985hydrodynamics, ZeyherFinger, FingerRice,widom2008discussion, landry2020effective}. In the continuum limit where the twisting angle or lattice mismatch is sufficiently small, the distinction between these cases is gone and the symmetry becomes a local continuous symmetry which gives rise to a true propagating Goldstone mode, the moir\'e phonon. We show how this discussion generalizes to multilayer systems with arbitrary twist angles leading to a simple counting rule for the number of phonons, moir\'e phonons, and phasons in the system. In particular, for a twisted $n$-layers system where all the successive twist angles between layers are small, non-zero, and have no special relation which makes their moir\'e patterns commensurate to one another, we find $2n$ gapless modes: two graphene phonons, two propagating moir\'e phonons and $2(n-2)$ diffusive phasons. We further studied twisted trilayer graphene with alternating twist angles and showed that its spectrum consists of a mirror-symmetric part that maps exactly to that of TBG at a rescaled twist angle $\sqrt{2/3} \theta$ as well as a mirror antisymmetric sector where all the modes are gapped.

One of our main conclusions is that, for TBG at small twist angles where the continuum approximation is expected to be valid, the phason perspective \cite{ochoa2019moire} should not lead to qualitatively new phenomena and the excitations are indistinguishable from propagating phonons. We stress here that this conclusion relies on a separation of scales between the graphene and moir\'e lattice scales i.e. $a/L_M \ll 1$ or equivalently small $\theta$. Thus, our expectation is that any correction to the results of the continuum model will vanish at least as $O(\theta)$ in terms of the natural energy scale of the problem $\omega_0$ (which itself scales with $\theta$). This means that as we deviate from the continuum limit and restore back the information about the underlying graphene lattice, we expect the moir\'e phonons to acquire a gap $\Delta$ and/or a damping term $\gamma$ that scales as $\theta$ for the commensurate (or incommensurate) case. Writing the general dispersion of the soft collective mode as
\begin{equation}\label{general_Phason}
    \omega^2 + i\gamma\omega = v_s^2 q^2 +\Delta^2
\end{equation}
Our discussion implies that: (i) for incommensurate stacking we have $\Delta = 0$ but finite $\gamma$, (ii) for commensurate stacking, we conclude that the gap and damping could be finite but bounded by the order of $\theta$ when $\theta$ is sufficiently small, and (iii) in the continuum limit, the distinction between the commensurate and incommensurate stacking is gone and we expect both $\gamma$ and $\Delta$ to vanish as $\gamma \sim \Delta \sim \theta\omega_0$. These results are summarized in Table~\ref{phenomenological_parameters}. We note that the appearance of a small gap for commensurate stacking away from the continuum limit was already observed in Ref.~\cite{maity2020phonons} which studied the lattice dynamics for twisted transition-metal dichalcogenide bilayers using atomistic classical simulations without any continuum approximation. For TBG at the magic angle, we would expect that the gap and the damping term (if nonzero) are of the order of $\sim 0.01$ meV, which should be hardly visible in experiments. 

Although the diffusive feature is suppressed in TBG at the magic angle, we do expect it to be measurable for large angles. In general, the diffusive nature of phason modes implies that they have a short lifetime (are overdamped). This means that they are not likely to contribute to experimental observables probing the low energy collective excitations at sufficiently low temperatures (compared to the scattering rate). For example, the diffusive nature of phasons can be detected through the measurement of specific heat at low temperatures~\cite{Cano2004explanantion,Baggioli2021new}. This is however beyond the current experimental capabilities in TBG due to the very small size of the samples. A more promising, albeit indirect way, is through the effect of electron-phonon coupling on the temperature dependence of resistivity \cite{maity2020phonons,Wu2019Phonon,Yudhistira2019Gauge,Ishizuka2021Purcell}. Finally, the damping can be inferred  from the measured phonon linewidth in the Raman spectra~\cite{JORIO20133,Barbosa_2022,c7010010}. In practice, all these estimates will be complicated by other sources of damping from e.g. disorder so that experimental measurements can only provide an upper bound for the effect.

\begin{table}[t]
\begin{tabular}{|l|ll|}
\hline
                & \multicolumn{1}{l|}{Commensurate} & \multicolumn{1}{l|}{Incommensurate} \\ \hline
\makecell[c]{Large $\theta$}         & \multicolumn{1}{l|}{\makecell[c]{$\Delta\neq 0$, $\gamma\neq 0$}  }          & \multicolumn{1}{l|}{\makecell[c]{$\Delta= 0$, $\gamma\neq 0$}    }          \\ \hline
\makecell[c]{Small $\theta$}          & \multicolumn{1}{l|}{\makecell[c]{$\Delta\sim \theta\omega_0$, $\gamma\sim \theta\omega_0$} }           & \multicolumn{1}{l|}{\makecell[c]{$\Delta=0$, $\gamma\sim \theta\omega_0$}   }           \\ \hline
Continuum Limit & \multicolumn{2}{l|}{\makecell[c]{$\Delta=0$, $\gamma=0$} }                                                  \\ \hline
\end{tabular}\caption{The dependence of the phenomenological parameters $\Delta$ and $\gamma$ for TBG on the twisting angle $\theta$, commensurability, and their continuum limit. Here $\omega_0$ is the natural energy scale of the system.}\label{phenomenological_parameters}
\end{table}

It is instructive to compare our results to those of Ref.~\cite{ochoa2022degradation} which deduced that phasons acquire a diffusive term as a result of anharmonic terms within the continuum limit. It is important to emphasize here that the damping term computed in that work is for the normal mode of the stacking variable $\delta {\boldsymbol \phi}$ rather than the soft mode variable $\bxi$. These two are not equivalent beyond the harmonic level due to the non-linear relationship between $\delta {\boldsymbol \phi}$ describing the soft mode and $\bxi$: $\delta {\boldsymbol \phi}(\br) = {\boldsymbol \phi}_0(\br + \bxi) - {\boldsymbol \phi}_0(\br)$, where ${\boldsymbol \phi}_0(\br)$ is the equilibrium configuration. Thus, while the calculation of Ref.~\cite{ochoa2022degradation} implies that the stacking variable (which may directly couple to some experimental probes) becomes overdamped, the collective lattice vibration modes described by the phason or moir\'e phonon remain undamped as we have shown in this work.

The same analysis can be applied to twisted multilayer graphene with alternating twist angles. Here, there is a single moir\'e pattern that gives rise to moir\'e phonons in the continuum limit very similar to the bilayer case. We note that in this setup, we do not get extra gapless modes upon increasing the number of layers since the moir\'e patterns generated by successive layers are perfectly aligned. Thus, we can think of this setup as a stack of two moir\'e that are perfectly commensurate with each other, leading to gapped modes even in the continuum limit with a gap $\Delta^{\text{ATTG}}_{\text{odd}} \sim \omega_0$ similar to what happens in AA or AB stacked graphene (note that this gap is different from the phason gap in Eq.~\eqref{general_Phason}). We have shown this explicitly for the trilayer case where the spectrum decomposes into a mirror-even that is identical to TBG spectrum (coming from the moir\'e pattern) and a mirror-odd part that corresponds to a commensurate AA stacking for the two moir\'e patterns between the two pairs of layers (12 and 32). One curious aspect is that similar to the electronic structure, the spectrum of excitations maps to the TBG case but with a different scaling for the angles: $\sqrt{2}$ for the electronic structure \cite{khalaf2019magic} and $\sqrt{3/2}$ for the lattice excitations. This suggests that at a given angle, the full system taking into account both electronic and lattice excitations is not exactly reducible to TBG (even when we focus on the mirror-even sector). On the other hand, we expect that whatever role played by moir\'e phonons in TBG, for example in explaining superconductivity\cite{wu2018theory,lian2019twisted} or transport \cite{Sharma2021}, will be qualitatively very similar for ATTG.

Finally, our results suggest a route to study the physics of phasons and their interaction with electrons by going beyond alternating twist configurations to multilayer systems with non-alternating twist angles realizing higher-order moir\'e pattern. Such systems represent a novel platform where the interplay between moir\'e-induced strong correlations and quasi-periodic structures can be explored. We leave a detailed study of this problem to future works.

\textit{Note Added:} Recently, we became aware of a recent study discussing the same gapless and gaped modes in ATTG and their coupling in the presence of an out-of-plane displacement field~\cite{Samajdar}.

\section*{Acknowledgement}
The authors acknowledge insightful discussions with Hector Ochoa and Rafael Fernandes regarding their related work. We are also grateful to Rhine Samajdar and Mathias Scheurer for informing us about their related work (posted simultaneously on the arXiv). Q.G. acknowledges the support of the Provost’s Graduate Excellence Fellowship from the University of Texas at
Austin. E.K. was supported by a Simons Investigator Award (PI: Ashvin Vishwanath). 

\appendix
\section{Isomorphism Between Mismatch and Phase Symmetry Groups and Equivalence Relation}\label{equivalenceRelation}

In this appendix, we elaborate on the implementation of the mismatch symmetry for incommensurate lattice and how it reduces to a phase symmetry.

The symmetry groups for the mismatch and phase symmetries in the two-chain model are given by
\begin{equation}
    \mathcal{G}_{\text{Pha}} = \{\epsilon=N_1 a-N_2 b,  |\Delta(\epsilon) = N_1a, \epsilon\in \mathcal{C}([0,b])\}
\end{equation}
where $\mathcal{C}([0,b])=\{ (Na \mod b) | N \in \mathbb{Z}\}$ is a covering set over the interval $[0,b]$ and becomes dense if $a/b \notin \mathbb{Q}$, and
\begin{equation}
    \mathcal{G}_{\text{Mis}} = \{N(a-b) |\Delta = Na, N\in \mathbb{Z}\}.
\end{equation}
Those two group are isomorphic to each other since $\mathcal{G}_{\text{Mis}} \cong \mathbb{Z} \cong \mathcal{G}_{\text{Pha}}$. 

One can also define an equivalence relation that two actions, shifting the top chain by $d_1$ and $d_2$, are equivalent to each other if they are the same module the lattice constant $b$: $ d_1\sim d_2: (d_1-d_2)\mod b =0$. This then implies that the mismatch group equals to the phase symmetry group: $\mathcal{G}_{\text{Pha}} = \mathcal{G}_{\text{Mis}}$ element-wise by noting that $\epsilon=N_1a-N_2b\sim N_1(a-b)$. Actually, this is to say that the group elements in $\mathcal{G}_{\text{Pha}} $ and $\mathcal{G}_{\text{Mis}}$ are identically equal to each other corresponding to the same symmetry actions. In the main text, we argue that those two symmetries are indeed equivalent to each other when acting on the Lagrangian. 

However, for $\mathcal{G}_{\text{Pha}}$ to represent a continuous symmetry when $\frac{a}{b} \notin \mathbb{Q}$, the group elements have to be arranged in a specific manner where $\epsilon$ varies continuously in $\mathcal{C}([0,b])$, which makes the shift of the matching point non-local, i.e. small $\epsilon$ will result in a large shift of the matching point. As the opposite, the group $\mathcal{G}_{\text{Mis}}$ represents a discrete but local symmetry if we arrange the group elements in the order of ascending $N$.

For the commensurate case where $\frac{a}{b} \in \mathbb{Q}$, the phase symmetry is degraded to a discrete symmetry but still non-local. Both the mismatch and phase symmetry will be isomorphic to a finite $\mathbb{Z}_n$: $ [\mathcal{G}_{\text{Mis}}](=\mathcal{G}_{\text{Mis}}/\sim)= \mathcal{G}_{\text{Pha}}\cong \mathbb{Z}_n$. The equivalence however is still exact. So, the mismatch symmetry is another representation of the phase symmetry, which is more convenient to be used in the continuum limit since it is local. 

\section{Correction to the Continuum Model}\label{correction_in_free_energy}
In this appendix, we estimate the error in Free energy of taking the continuum limit for the stacking two-chain model, which can be applied to TBG through the correspondences in Table~\ref{table_correspondences}.

The Free energy of the two chains can be written as
\begin{equation}\label{DiscreteFE}
    F = U_{El} + U_B
\end{equation}
where
\begin{equation}
\begin{split}
    U_{El} =& \sum_{l}\left[\frac{\kappa/b}{2}(u^{(2)}_{(l+1)b} - u^{(2)}_{lb})^2 + \frac{\kappa/a}{2}(u^{(1)}_{(l+1)a} - u^{(1)}_{la})^2\right] \\
    \equiv& U^{(2)}_{El} + U^{(1)}_{El}
\end{split}
\end{equation}
is the elastic energy of the two chains
and
\begin{equation}\label{bindingEnergyInLattice}
    U_B = -\sum_{l} bV_0\cos\frac{2\pi}{a}\left[\left(1-\frac{a}{b}\right)lb + (u^{(2)}_{lb} - u^{(1)}_{[lb]})\right],
\end{equation}
is the binding energy between the two chains.
Here $\kappa$ is the elastic constant (with unit of energy per unit cell), $V_0$ is the inter-chain coupling constant (with unit of energy per unit cell of chain 2) with a minus sign in front of it meaning that the ``AA" stacking is preferred. $u^{(2)}_{lb}$ is the displacement of the $l^{\text{th}}$ atom at chain 2 and $u^{(1)}_{[lb]}$ is the displacement of the atom at chain 1 that is closest to the $l^{\text{th}}$ atom at chain 2.

In the continuum model where $1-a/b \ll 1$, by taking $lb,[lb]\to x$ and $a(\approx b)/L_M \to 0$ with $L_M = a/(1-a/b) \gg a$, the Free energy can be expressed as~\cite{nam2017lattice}:
\begin{equation}\label{continuum_two_chain}
\begin{split}
    \tilde F =& \int dx [U_{El}(x) + U_B(x)] \\
    =& \int dx \sum_{\alpha=1,2}\frac{\kappa}{2}(\partial_x u^{(\alpha)}(x))^2 \\
    &- \int dx V_0\cos\left[G_M x +\frac{2\pi}{a} u^{(21)}(x)\right]
\end{split}
\end{equation}
where $G_M = 2\pi/L_M$.
We can rewrite $\tilde F $ in a discrete form:
\begin{equation}
    \begin{split}
        \tilde F =& \sum_{l} \int_{lb}^{(l+1)b} dx [U^{(2)}_{El}(x) + U_B(x)] \\
        &+ \sum_{l} \int_{la}^{(l+1)a} dx U^{(1)}_{El}(x) \\
        =& \sum_{l} \int_{lb}^{(l+1)b} dx [U^{(2)}_{El}(x=lb) + U_B(x=lb)] \\
        &+ \sum_{l} \int_{la}^{(l+1)a} dx U^{(1)}_{El}(x=la) \\
        &+ \sum_{l} \int_{lb}^{(l+1)b} dx (x-lb) \partial_x[U^{(2)}_{El}(x=lb) + U_B(x=lb)] \\
        &+ \sum_{l} \int_{la}^{(l+1)a} dx (x-la) \partial_x U^{(1)}_{El}(x=la)+ \cdots,
    \end{split}
\end{equation}
where one can see that the first two terms are (or close to) the discrete Free energy in Eq.~\eqref{DiscreteFE} while the rest are then extra terms induced by taking the continuum limit. We note here that we overlooked the mismatch between $a$ and $b$, i.e., we took $[lb]$ to be $lb$, which will introduce an error of order $O(a/L_M)$ or $O(b/L_M)$. 
Thus the correction from the continuum model to the original lattice model can be estimated by
\begin{equation}\label{correction0}
    \delta F = \tilde F - F \approx \frac{L}{L_M}\int_0^{L_M} dx \frac{b}{2}\partial_x[U_{El}(x) + U_B(x)],
\end{equation}
where we assumed that $a\approx b$ and $L$ is the length of the whole system (which should be fixed). 
Since $U_{El}(x) $ contains only derivatives of $u^{(l)}(x)$, it will contribute higher order correction to the $ \delta F$. Thus, the leading order comes from
\begin{equation}\label{correction}
    \begin{split}
        \delta F \sim& \frac{L}{L_M}\int_0^{L_M} dx \frac{b}{2}\partial_x U_B(x) \\
        = & \frac{L}{L_M}\int_0^{L_M} dx \frac{b}{2}\left[\frac{2\pi}{L_M}+ \frac{2\pi}{a}\partial_x u^{(21)}(x)\right]\\
        &\times V_0\sin\left[G_M x +\frac{2\pi}{a} u^{(21)}(x)\right].
    \end{split}
\end{equation}
The key to the scaling behavior of $\delta F$ is the variation of $ u^{(21)}(x)$ over space. It can be shown that for very large $L_M$, domain walls are formed as the equilibrium configuration corresponding to Eq.~\eqref{continuum_two_chain} is a soliton~\cite{nam2017lattice}, and the profile of the relative displacement can be approximated by
\begin{equation}
    u^{(21)}(x)\sim \,\,\begin{cases}
	ax/(L_M-w_d)&		0<x<L_M-w_d\\
	c_0-a(x-d_0)/w_d&		L_M-w_d<x<L_M,\\
\end{cases}
\end{equation}
where $w_d$ is the width of the domain wall and assumed to be much smaller than the moir\'e length scale $w_d \ll L_M$, and $c_0$, $d_0$ are constants ensuring that the profile is continuous. The change of the displacement $u^{(21)}(x) $ near the connecting point $x=L_M-w_d$ is actually smooth which can be confirmed numerically, so we can ignore the contribution from it.  Plugging the profile into Eq.~\eqref{correction}, we have
\begin{equation}
    \begin{split}
        &\delta F \sim \frac{L}{L_M}\int_0^{L_M-w_d} dx \frac{2\pi b}{L_M-w_d} V_0\sin\left[G_M x +\frac{2\pi}{a} u^{(21)}(x)\right]\\
        &- \frac{L}{L_M}\int_{L_M-w_d}^{L_M} dx \frac{2\pi b}{w_d} V_0\sin\left[G_M x +\frac{2\pi}{a} u^{(21)}(x)\right] \\
        &=\frac{b}{L_M}L\left[\frac{1}{L_M-w_d}\int_0^{L_M-w_d} - \frac{1}{w_d}\int_{L_M-w_d}^{L_M}\right] \\
        &\quad\times dx V_0\sin\left[G_M x +\frac{2\pi}{a} u^{(21)}(x)\right]\\
        &\sim O(b/L_M)F
    \end{split}
\end{equation}
where we have assumed that the integration over $V_0 \sin\left[G_M x +\frac{2\pi}{a} u^{(21)}(x)\right]$ gives roughly the same energy scale ($\sim LV_0$) as the original free energy. Thus, we have that the energy correction is approximately:
\begin{equation}
    \frac{\delta F}{F} \sim O(b/L_M),
\end{equation}
which simply means that what has been omitted during the continuum approximation is at the order of $O(\theta)$ with respect to what has been kept. 
The same story can be applied to TBG where the 1D integral becomes 2D (the domain walls change from segments in 1D to stripes in 2D), but the scaling is the same that $\delta F^{\text{TBG}}/F^{\text{TBG}}\sim O(\theta)$.

\section{Variable Transformations}\label{variable_transformations}
The Lagrangian for the vibration mode $\delta \bu(\br,t)$ that gives rise to the equation of motion in Eq.~\eqref{ELDyn} reads
\begin{equation}\label{lagrangianDensity}
\begin{split}
    \mathcal{L}[\delta \bu,\partial\delta \bu] =&\frac{\rho}{4}(\delta\dot{\bu})^2- \frac{1}{2}\left[ \frac{\lambda}{2}(\boldsymbol{\nabla}\cdot\delta\bu)^2 +\frac{\mu}{4}(\partial_i \delta u_{j}+\partial_j \delta u_{i})^2 \right.\\
    &\left.+ \delta u_{i}\delta u_{j}\frac{\partial^2 V^{(21)}}{\partial u_i^{(21)}\partial u_j^{(21)}}\bigg|_{\bu^{(21)}_0(\boldsymbol{r})} \right].
\end{split}
\end{equation}
We have the transformation between the vibration $\delta \bu$ and the symmetry generator $\bxi$:
\begin{equation}
    \delta u_i = \xi_j\partial_j u_i^{(21)} + (1- \mathcal{R}^{-1}_{\theta})_{ij}\xi_j \equiv \mathcal{C}_{ij}\xi_j.
\end{equation}
Then, it is straightforward to see that in the Lagrangian, the terms transform in the following ways:
\begin{equation}
    \begin{split}
        \boldsymbol{\nabla}\cdot\delta\bu = \partial_i\delta u_i &= \partial_i \left[ \xi_j\partial_j u_i^{(21)} + (1- \mathcal{R}^{-1}_{\theta})_{ij}\xi_j \right] \\
        &= \mathcal{C}_{ij}\partial_i\xi_j + \xi_j\partial_j\partial_i u_i^{(21)},
    \end{split}
\end{equation}
\begin{widetext}
\begin{equation}
    \begin{split}
        \Rightarrow (\boldsymbol{\nabla}\cdot\delta\bu)^2 &= (\mathcal{C}_{ij}\partial_i\xi_j + \xi_j\partial_j\partial_i u_i^{(21)})(\mathcal{C}_{lm}\partial_l\xi_m + \xi_m\partial_m\partial_l u_l^{(21)}) \\
        &=\mathcal{C}_{ij}\mathcal{C}_{lm}\partial_i\xi_j\partial_l\xi_m + 2\mathcal{C}_{ij}\partial_m\partial_l u_l^{(21)}\xi_m\partial_i\xi_j  + \partial_j\partial_i u_i^{(21)}\partial_m\partial_l u_l^{(21)}\xi_m\xi_j,
    \end{split}
\end{equation}
\begin{equation}
    \begin{split}
        \partial_i \delta u_{j}+\partial_j \delta u_{i} &= \partial_i \left[ \xi_k\partial_k u_j^{(21)} + (1- \mathcal{R}^{-1}_{\theta})_{jk}\xi_k \right] + \partial_j \left[ \xi_k\partial_k u_i^{(21)} + (1- \mathcal{R}^{-1}_{\theta})_{ik}\xi_k \right] \\
        &= \mathcal{C}_{jk}\partial_i\xi_k + \mathcal{C}_{ik}\partial_j\xi_k +  \xi_k\partial_k(\partial_i u_j^{(21)} + \partial_j u_i^{(21)}),
    \end{split}
\end{equation}
\begin{equation}
    \begin{split}
        \Rightarrow (\partial_i \delta u_{j}+\partial_j \delta u_{i} )^2=& \left[ \mathcal{C}_{jm}\partial_i\xi_m + \mathcal{C}_{im}\partial_j\xi_m +  \xi_m\partial_m(\partial_i u_j^{(21)} + \partial_j u_i^{(21)}) \right]\left[ \mathcal{C}_{jl}\partial_i\xi_l + \mathcal{C}_{il}\partial_j\xi_l +  \xi_l\partial_l(\partial_i u_j^{(21)} + \partial_j u_i^{(21)}) \right] \\
        =& 2\mathcal{C}_{jm}\mathcal{C}_{jl}\partial_i\xi_m\partial_i\xi_l + 2\mathcal{C}_{jm}\mathcal{C}_{il}\partial_i\xi_m\partial_j\xi_l + 4\mathcal{C}_{jm}\partial_l(\partial_i u_j^{(21)} + \partial_j u_i^{(21)})\xi_l\partial_i\xi_m \\
        &+ \partial_m(\partial_i u_j^{(21)} + \partial_j u_i^{(21)})\partial_l(\partial_i u_j^{(21)} + \partial_j u_i^{(21)})\xi_m\xi_l
    \end{split}
\end{equation}
and
\begin{equation}
    \begin{split}
        \delta u_{i}\delta u_{j}\frac{\partial^2 V^{(21)}}{\partial u_i^{(21)}\partial u_j^{(21)}}\bigg|_{\bu^{(21)}_0(\boldsymbol{r})} =& \xi_l\xi_m\partial_l\left[u_i^{(21)} + (1- \mathcal{R}^{-1}_{\theta})_{ik} r_k\right]\partial_m \left[u_j^{(21)} + (1- \mathcal{R}^{-1}_{\theta})_{jk} r_k\right] \frac{\partial^2 V^{(21)}}{\partial u_i^{(21)}\partial u_j^{(21)}}\bigg|_{\bu^{(21)}_0(\boldsymbol{r})}\\
        =& \xi_l\xi_m\mathcal{C}_{il}\partial_m \frac{\partial V^{(21)}}{\partial u_i^{(21)}}\bigg|_{\bu^{(21)}_0(\boldsymbol{r})}.
    \end{split}
\end{equation}
Plugging these terms into the Lagrangian, we have $ \mathcal{L}[ \bxi,\partial\bxi] = \mathcal{T}[\bxi,\partial\bxi]-\mathcal{V}[\bxi,\partial\bxi]$ with the potential term
\begin{equation}\label{pot}
\begin{split}
    &\mathcal{V}[\bxi,\partial\bxi] =  \frac{1}{2}\left\{\frac{1}{2}\left[ \lambda\mathcal{C}_{ij}\mathcal{C}_{lm} + \mu\mathcal{C}_{kj}\mathcal{C}_{km}\delta_{il}+\mu\mathcal{C}_{im}\mathcal{C}_{lj} \right]\partial_i\xi_j\partial_l\xi_m + \left[ \lambda\mathcal{C}_{ij}\partial_m\partial_l u^{(21)}_l+\mu \mathcal{C}_{lj}\partial_m(\partial_iu^{(21)}_l+\partial_l u^{(21)}_i) \right]\xi_m\partial_i\xi_j \right. \\
    &+ \left.\left[ \frac{\lambda}{2}\partial_i(\partial_l u^{(21)}_l)\partial_j(\partial_m u^{(21)}_m) + \frac{\mu}{4}\partial_i(\partial_l u^{(21)}_m+\partial_m u^{(21)}_l)\partial_j(\partial_l u^{(21)}_m+\partial_m u^{(21)}_l)+\mathcal{C}_{ki}\partial_j \frac{\partial V^{(21)}}{\partial u_k^{(21)}}\bigg|_{\bu^{(21)}_0(\boldsymbol{r})} \right]\xi_i\xi_j\right\}\\
    &\equiv \frac{1}{2}(\mathcal{A}_{ijlm}\partial_i\xi_j\partial_l\xi_m + \mathcal{B}_{ijm}\xi_m\partial_i\xi_j + \mathcal{D}_{ij}\xi_i\xi_j),
\end{split}
\end{equation}
\end{widetext}
and the generator $\bxi$, and the kinetic term:
\begin{equation}\label{kinetic}
    \mathcal{T}[\bxi,\partial\bxi] = \frac{\rho}{4}\mathcal{C}_{li}\mathcal{C}_{lj}\dot\xi_i\dot\xi_j\equiv\frac{1}{2}\mathcal{M}_{ij}\dot\xi_i\dot\xi_j.
\end{equation}
One can show the following identities:
\begin{equation}
    \begin{split}
        \mathcal{A}_{ijlm} &= \mathcal{A}_{lmij} , \\
        \mathcal{D}_{ij} &= \mathcal{D}_{ji}, \\
        \partial_i \mathcal{B}_{ijm} &= 2\mathcal{D}_{jm}. \\
    \end{split}
\end{equation}
Thus, we can define $\mathcal{B}_{ijm}^{\pm}\equiv\frac{1}{2}( \mathcal{B}_{ijm}\pm\mathcal{B}_{imj} )$, and we will have:
\begin{equation}
    \begin{split}
        \partial_i \mathcal{B}_{ijm}^{+} &= 2\mathcal{D}_{jm}, \\
        \partial_i \mathcal{B}_{ijm}^{-} &= 0.
    \end{split}
\end{equation}
Using $\mathcal{B}_{ijm}^{\pm}$, we can rewrite the Lagrangian as

\begin{equation}\label{newLagrangian}
    \mathcal{L}[ \bxi,\partial\bxi] = \frac{1}{2}\mathcal{M}_{ij}\dot\xi_i\dot\xi_j - \frac{1}{2}(\mathcal{A}_{ijlm}\partial_i\xi_j\partial_l\xi_m + \mathcal{B}_{ijm}^{-}\xi_m\partial_i\xi_j)
\end{equation}
by noting that
\begin{equation}
\begin{split}
    &\mathcal{B}_{ijm}^{+}\xi_m\partial_i\xi_j \\
    &= \partial_i(\mathcal{B}_{ijm}\xi_m\xi_j)-(\partial_i\mathcal{B}_{ijm}^{+})\xi_m\xi_j - \mathcal{B}_{ijm}^{+}\xi_j\partial_i\xi_m \\
    &= \partial_i(\mathcal{B}_{ijm}\xi_m\xi_j)-(\partial_i\mathcal{B}_{ijm}^{+})\xi_m\xi_j - \mathcal{B}_{ijm}^{+}\xi_m\partial_i\xi_j,
\end{split}
\end{equation}
which leads to 
\begin{equation}
    \mathcal{B}_{ijm}^{+}\xi_m\partial_i\xi_j = \frac{1}{2}\partial_i(\mathcal{B}_{ijm}\xi_m\xi_j)-\frac{1}{2}(\partial_i\mathcal{B}_{ijm}^{+})\xi_m\xi_j.
\end{equation}
It is worth noting that $\mathcal{B}_{ijm}^{-}$ is found to be a total derivative of the displacement $\bu$:
\begin{equation}
    \mathcal{B}_{ijm}^{-}=\epsilon_{jmn}\epsilon_{npq}\partial_p \tilde B_{iq}
\end{equation}
with
\begin{equation}
    \tilde B_{iq} \equiv \frac{1}{2}[\lambda\mathcal{C}_{iq}\partial_l u_l + \mu \mathcal{C}_{lq}(\partial_i u_l+\partial_l u_i)],
\end{equation}
where $\epsilon_{jmn}$ is the Levi-Civita symbol.

\section{Low Energy Effective Field Theory for Lattice Dynamics in TBG}\label{EffFieldTheory}
We start with the action for the lattice dynamics in TBG which is written as
\begin{equation}\label{Seff}
\begin{split}
    S =& \int d^2 \br dt \L = \frac{1}{2} \int d^2 \br dt \\
    &\times \left\{ \dot \xi_i \M_{ij} \dot \xi_j - \B_{ijm}^- \xi_m \partial_i \xi_j - \A_{ijlm} \partial_i \xi_j \partial_l \xi_m \right\}
\end{split}
\end{equation}
It is more convenient to work under the Fourier basis:
\begin{equation}
\begin{split}
    \xi_i(\br) &= \int d^2 \bq d\omega e^{i (\bq \cdot \br - \omega t)} \xi_i(\bq, \omega),\\
    \M_{ij}(\br) &= \sum_\bG e^{i \bG \cdot \br} \M_{ij}(\bG),\\
    \B_{ijm}(\br) &= \sum_\bG e^{i \bG \cdot \br} \B_{ijm}(\bG),\\
    \A_{ijlm}(\br) &= \sum_\bG e^{i \bG \cdot \br} \A_{ijlm}(\bG),
\end{split}
\end{equation}
where the expansions for $\M$, $\A$, and $\B$ contain only reciprocal lattice vectors $\bG$ since they are periodic whereas the $\bq$ integral in the Fourier expansion of $\xi$ is over all momenta. Substituting in Eq.~\eqref{Seff} yields
\begin{equation}\label{Sq}
\begin{split}
    S =& \frac{1}{2} \int d^2 \bq d\omega \\
    &\times\sum_\bG \xi_i(\bq, \omega) \Gamma_{ij}(\bq, \bq + \bG, \omega) \xi_j(-\bq - \bG, -\omega) 
\end{split}
\end{equation}
with
\begin{equation}
    \begin{split}
        \Gamma_{ij}(\bq, \bq + \bG)=& \omega^2 \M_{ij}(\bG)- q_l (q_k + G_k) \A_{likj}(\bG) \\
        &  - i q_m \B^-_{mij}(\bG)
    \end{split}
\end{equation}
where we used the relation $\partial_m \B^-_{mij}=0$ which implies $G_m \B^-_{mij} = 0$.
Naively, we derive an effective field theory by focusing on small momenta $\bq$ which amounts to setting $\xi_{\bq}$ to zero for $|\bq|$ larger than a certain cutoff (which is a lot smaller than any non-zero reciprocal lattice vector $\bG \neq 0$). This amounts to taking the $\bG = 0$ term in the sum above which is equivalent to averaging $\M_{ij}(\br)$, $\B_{ijm}(\br)$, and $\A_{ijlm}(\br)$ in Eq.~\eqref{Seff} over the unit cell. This is however incorrect since the action in Eq.~\eqref{Sq} contains terms that couple the slowly varying modes $\xi_\bq$ with the fast modes $\xi_{\bq + \bG}$ which means that the proper procedure is to integrate out the fast modes which will yield important contributions to the action. To perform the integral we write $\bq = \bk + \bG$ where $\bk$ is in the first BZ and restrict ourselves to small $|\bk| < \Lambda$ (since different momenta within the BZ are not coupled). We then obtain (we will drop the omega dependence which will be left implicit)
\begin{equation}
\begin{split}
    S =& \frac{1}{2} \int_{|\bk| < \Lambda} d^2 \bk \\ &\times\sum_{\bG,\bG'} \xi_i(\bk + \bG) \xi_j(-\bk - \bG') \Gamma_{ij}(\bk + \bG, \bk + \bG') \\
    =& \frac{1}{2} \int_{|\bk| < \Lambda} d^2 \bk \Bigg\{ \xi_i(\bk) \Gamma_{ij}(\bk, \bk) \xi_j(-\bk)\\
    &\qquad + \sum_{\bG \neq 0} \xi_i(\bk + \bG)  \Gamma_{ij}(\bk + \bG, \bk) \xi_j(- \bk) \\
    &\qquad  + \sum_{\bG' \neq 0} \xi_i(\bk)  \Gamma_{ij}(\bk, \bk + \bG') \xi_j(- \bk - \bG') \\ 
     & \qquad   + \sum_{\bG, \bG' \neq 0} \xi_i(\bk + \bG)  \Gamma_{ij}(\bG, \bG') \xi_j(-\bk - \bG') \Bigg\}
\end{split}
\end{equation}
Here, we have split the sum into four terms according to whether $\bG$ or $\bG'$ is zero. We have also used the fact that $|\bk| \ll |\bG|$ whenever $\bG \neq 0$ to set $\bk$ to zero whenever it is added to non-zero $\bG$ in $\Gamma$. Note that the first term is the term that we get by averaging the prefactors over a unit cell. The extra terms contain fast varying fields. We notice that $\omega$ is of the same order as $\bk$ so it can be neglected in the second and third terms. The Gaussian integral over the fast variable $\xi(\bk + \bG)$ can be readily performed and yielding
\begin{equation}
    S = \frac{1}{2} \int_{|\bk| < \Lambda} d^2 \bk  \xi_i(\bk) \xi_j(-\bk) \tilde \Gamma_{ij}(\bk, \bk) 
\end{equation}
with
\begin{gather}
\tilde \Gamma_{ij}(\bk, \bk) = \Gamma_{ij}(\bk, \bk) - \Delta \Gamma_{ij}(\bk, \bk) \nonumber \\
    \Delta \Gamma_{ij}(\bk,\bk) =  \sum_{\bG, \bG' \neq 0}\Gamma_{il}(\bk, \bk + \bG) \Gamma^{-1}_{lm}
    (\bG, \bG') \Gamma_{mj}(\bk + \bG', \bk)
    \label{DeltaGamma}
\end{gather}
where the second term schematically has the form $\sim \xi(\bk) \Gamma(\bk,\bG) \Gamma^{-1}(\bG,\bG') \Gamma(\bG',\bk) \xi(-\bk)$. Such term is of order $\bk^2$ (since $\Gamma(\bG,\bk) \sim O(\bk)$ and $\Gamma(\bG,\bG') \sim O(1)$) and thus contributes to the velocity.

To evaluate the action to leading order in $\bk \sim \omega \sim \epsilon$, we keep terms of order $O(\epsilon^0)$ in $\Gamma_{ij}(\bG, \bG')$, terms of order $O(\epsilon^1)$ in $\Gamma_{ij}(\bk, \bk + \bG)$ and terms of order $O(\epsilon^1)$ in $\Gamma_{ij}(\bk, \bk + \bG)$ giving
\begin{equation}
\begin{split}
    \Gamma_{ij}(\bG, \bG') =& - G_l G'_k \A_{likj}(\bG' - \bG) \\
    &- \frac{i}{2} (G_m + G'_m) \B^-_{mij}(\bG' - \bG) \\
    \Gamma_{ij}(\bk, \bk + \bG) =& - k_l G_k \A_{likj}(\bG) - i k_m \B^-_{mij}(\bG) \\
    \Gamma_{ij}(\bk + \bG, \bk) =& - G_l k_k \A_{likj}(-\bG) - i k_m \B^-_{mij}(-\bG) \\
    \Gamma_{ij}(\bk,\bk) =& \omega^2 \M_{ij}(0) - k_l k_k \A_{likj}(0)
    \label{GammaDef}
\end{split}
\end{equation}
where we have used the fact that $\B^-_{ijm}(0) = 0$. If we recast the variables from momentum space to real space, we shall have the Lagrangian for the low-energy effective field theory:
\begin{equation}
    \mathcal{L}_{\text{eff}} = \frac{1}{2} \M_{ij}(0) \dot \xi_i  \dot \xi_j - \frac{1}{2} \tilde \A_{ijlm} \partial_i \xi_j \partial_l \xi_m,
\end{equation}
where
\begin{equation}
    \tilde \A_{ijlm} = \A_{ijlm}(0) + \Delta_{ijlm}
\end{equation}
with
\begin{equation}\label{selfEnergy}
    \begin{split}
        \Delta_{ijlm} \equiv&\sum_{\bG,\bG'\neq 0}[iG_k\A_{ijkp}(\bG)-\B^-_{ijp}(\bG)]\Gamma^{-1}_{pq}(\bG, \bG')\\
        &\qquad\times[-iG_n\A_{nqlm}(-\bG)+\B^-_{lqm}(-\bG)]
    \end{split}
\end{equation}
giving the self-energy correction due to the coupling between slow varying and fast varying modes.

\section{Small $\theta$ limit}\label{smallTheta}
In this appendix, we derive the effective theory parameters in the small angle limit Eq.~\eqref{SmallthetaFTP}.

The Euler Lagrange equations (given by Eq.~\eqref{EquPosBi} in the main text and repeated here for completeness) has the form
\begin{gather}
    (\lambda + \mu)\boldsymbol{\nabla}(\boldsymbol{\nabla}\cdot \bphi) + \mu\boldsymbol{\nabla}^2\bphi = 2\frac{\partial V^{(21)}}{\partial \bphi}, \nonumber \\ {\boldsymbol \phi}(\br) = \bu^{(21)}(\br) + (1 - \R_\theta^{-1}) \br
    \label{EL}
\end{gather}
The potential $V^{(21)}$ is periodic under $\bphi(\br) = \bphi(\br) + \ba_l$, $l=1,2$ where $\ba_l$ are the graphene lattice vectors which we choose to be $\ba_1 = (1,0) a$ and $\ba_2 = (1/2, \sqrt{3}/2) a$ where $a$ is the graphene lattice constant $a = \sqrt{3} a_{CC} \approx 2.46 A^o$. In the following, we will set $a$ to 1 and only restore it in the final expression. The potential has it minima at the AB and BA stacking points corresponding to $\bphi_{\rm AB} = (\frac{1}{2}, -\frac{1}{2\sqrt{3}})$ and $\bphi_{\rm BA} = (\frac{1}{2}, \frac{1}{2\sqrt{3}})$ and those related by $R_{2\pi/3}$. We are looking for solutions of Eq.~\eqref{EL} where $\bu^{(21)}(\br)$ is periodic under moir\'e translations $\bL_l = [1 - \R_{-\theta}]^{-1} \ba_l$. This implies
\begin{equation}
    \bphi(\br + \bL_l) = \bphi(\br) + (1 - \R_{-\theta}) \bL_l =  \bphi(\br) + \ba_l
    \label{BoundaryMoire}
\end{equation}
Thus, $\bphi$ to shift by a lattice vector upon shifting $\br$ by a moir\'e lattice vector.

We will now assume that the angle $\theta$ is very small and see how we can simplify Eq.~\eqref{EL}. First, notice that $\theta$ only enters in the boundary condition (\ref{BoundaryMoire}). If we first neglect the boundary conditions, we see that choosing $\bphi$ to be a constant and equal to the value which minimizes (or maximizes) the potential yields a solution to Eq.~\eqref{EL}. However, a constant $\bphi$ does not satisfy the boundary condition (\ref{BoundaryMoire}).  Notice that if $\bphi$ changes over length scale $\kappa \lesssim L_M$, then the LHS is of order $(\mu/\kappa^2) \bphi$. On the other hand, the RHS side away from the extrema is of the order $V_0$. This introduces the length scale $w \sim \sqrt{\mu/V_0}$ such that at small angles $w \ll L_M$ and the change of $\bphi$ is governed by $w$. The resulting solution is one where $\bphi$ consists of domains where it is a constant equal to $\bphi_{\rm BA}$ or $\bphi_{\rm AB}$ separated by domain walls with thickness $\sim w$ \cite{koshino2019moire, ochoa2019moire}. Let us now consider a domain wall along the $y$-direction connecting $\bphi_{\rm AB}$ and $\bphi_{\rm BA}$ such that
\begin{gather}
    \bphi(x = -\infty) = \phi_{\rm AB} = (1/2, -1/(2\sqrt{3})), \nonumber \\ \bphi(x = +\infty) = \phi_{\rm BA} = (1/2, 1/(2\sqrt{3})), 
    \label{BCDW}
\end{gather}
We notice that neither Eq.~\eqref{EL} nor Eq.~\eqref{BCDW} depend explicitly on $y$ so we can look for solutions which only depend on $x$: $\bphi(\br) = \bphi(x)$. Substituting in Eq.~\eqref{EL} we get
\begin{gather}
    (\lambda + 2\mu) \phi_x'' = -4\pi V_0 \sin 2\pi \phi_x \cos \frac{2\pi}{\sqrt{3}} \phi_y ,\\
    \mu \phi_y'' = -\frac{4 \pi}{\sqrt{3}} V_0 (\cos 2\pi \phi_x + 2 \cos \frac{2\pi}{\sqrt{3}} \phi_y) \sin \frac{2\pi}{\sqrt{3}} \phi_y.
\end{gather}
We see that $\phi_x = 1/2$ satisfies the first equation as well as the boundary condition for $\phi_x$. We are then left with the equation
\begin{align}
    \mu \phi_y'' &= -\frac{4 \pi}{\sqrt{3}} V_0 (-1 + 2 \cos \frac{2\pi}{\sqrt{3}} \phi_y) \sin \frac{2\pi}{\sqrt{3}} \phi_y, \nonumber \\
    &= \frac{4 \pi}{\sqrt{3}} V_0 (\sin \frac{2\pi}{\sqrt{3}} \phi_y - \sin \frac{4\pi}{\sqrt{3}} \phi_y).
\end{align}
Defining $f(x) = \frac{2\pi}{\sqrt{3}} \phi_y(x)$, we can write this as
\begin{equation}
    f'' = w^{-2} \left(\sin f - \sin 2f \right), \qquad f(x = \pm \infty) = \pm \frac{\pi}{3},
\end{equation}
where $w$ is defined via
\begin{equation}
    w =  a\sqrt{\frac{3\sqrt{3}}{8\pi^2} \frac{\mu}{V_0}}.
\end{equation}
The equation above can be solved using standard methods by first multiplying both sides by $f'$, writing the LHS as $\frac{d}{dx} f'^2$ and integrating twice leading to
\begin{equation}
    f(x) = 2 \arctan \left(\frac{1}{\sqrt{3}} \tanh \sqrt{\frac{3}{2}} \frac{x}{2 w} \right).
\end{equation}
The general solution consists of a sum of three domain wall solutions
\begin{equation}
    \bphi_0(\br) = \sum_{l=0}^2 \R_{2\pi l/3} \bphi^{\rm DW}(\R_{2\pi l/3} \br),
\end{equation}
where $\phi^{\rm DW}(\br) = g(x) = \frac{\sqrt{3}}{2\pi} f(x)$.

To derive the effective field theory parameters, we start with the function $\C_{ij} = \partial_j \phi_{0,i}(\br)$. For the domain wall parallel to the $y$-axis, we get
\begin{equation}
    \C^{\rm DW}_{ij}(\br) = \delta_{i,y} \delta_{j,x} g'(x).
\end{equation}
The full function $\C_{ij}$ is generated from $\C^{\rm DW}$ by rotation
\begin{equation}
    \C_{ij}(\br) = \sum_{l=0}^2 [\R_{2\pi l/r}]_{ii'} [\R_{2\pi l/r}]_{jj'} \C^{\rm DW}_{i'j'}(\R_{2\pi l/3} \br).
\end{equation}
Next we use the fact that at small $\theta$ we can ignore the regions of overlap of domain walls to write $\A$ and $\B$ defined in Eq.~\eqref{pot} similarly in terms of a single domain wall solution
\begin{gather}
    \A_{ijkm}(\br) = \sum_{l=0}^2 [\R_{2\pi l/r}]_{ii'} [\R_{2\pi l/r}]_{jj'} [\R_{2\pi l/r}]_{kk'} \nonumber \\ \times  [\R_{2\pi l/r}]_{mm'} \A^{\rm DW}_{i'j'k'm'}(\R_{2\pi l/3} \br), \nonumber \\ \B_{ijm}(\br) = \sum_{l=0}^2 [\R_{2\pi l/r}]_{ii'} [\R_{2\pi l/r}]_{jj'}  [\R_{2\pi l/r}]_{mm'} \nonumber \\ \times \B^{\rm DW}_{i'j'm'}(\R_{2\pi l/3} \br), \nonumber \\ \M_{ij}(\br) = \sum_{l=0}^2 [\R_{2\pi l/r}]_{ii'} [\R_{2\pi l/r}]_{jj'} \M^{\rm DW}_{i'j'}(\R_{2\pi l/3} \br).
\end{gather}
We can now use the simple form of $\C^{\rm DW}$ to write simple expressions for $\M^{\rm DW}$, $\B^{\rm DW}$ and $\A^{\rm DW}$:
\begin{gather}
    \M^{\rm DW}_{ij}(\br) = \delta_{i,x} \delta_{j,x} g'(x)^2,\nonumber \\
    \B^{\rm DW}_{ijm}(\br) = \mu \delta_{i,x} \delta_{j,x} \delta_{m,x} g'(x) g''(x), \nonumber \\
    \A^{\rm DW}_{ijlm}(\br) = \delta_{j,x} \delta_{m,x}[ \mu \delta_{i,x} \delta_{l,x} + (\lambda + 2\mu) \delta_{i,y} \delta_{l,y}] g'(x)^2.
    \label{DWMatrices}
\end{gather}

The main remaining component to compute the field theory parameters is the corrections from integrating out massive modes given in Eq.~\eqref{DeltaGamma}. This can be simplified by introducing
\begin{equation}
    \psi_{lj}(\bG, \bk) = \sum_{\bG' \neq 0} \Gamma_{lm}^{-1}(\bG, \bG') \Gamma_{mj} (\bk + \bG', \bk),
\end{equation}
which enables us to write
which can be written as
\begin{align}
    \Delta \Gamma_{il}(\bk) &=  \sum_\bG \Gamma_{ij}(\bk, \bk + \bG) \psi_{jl}(\bG, \bk) \nonumber \\
    &= -k_m \sum_\bG (G_k \A_{mikj}(\bG) + i \B^-_{mij}(\bG)) \psi_{jl}(\bG, \bk) \nonumber  \\&= ik_m \int \frac{d^2 \br}{A_M}  \psi_{jl}(-\br, \bk) ( \partial_k \A_{mikj}(\br)  - \B^-_{mij}(\br) ) .
    \label{DeltaGammaPsi}
\end{align}
Now note that $\psi$ is a solution to the equation
\begin{equation}
    \sum_{\bG'} \Gamma_{ij}(\bG, \bG') \psi_{jp} (\bG', \bk) = \Gamma_{ip}(\bG + \bk, \bk).
\end{equation}
Here we have extended the summation to include $\bG' = 0$ by choosing $\psi_{jp} (\bG', \bk) = 0$. The LHS of this equation can be simplified as
\begin{widetext}
\begin{align}
    \sum_{\bG'} \Gamma_{ij}(\bG, \bG') \psi_{jp} (\bG', \bk) &= \sum_{\bG'} \int \frac{d^2 \br}{A_M} e^{-i (\bG' - \bG) \cdot \br} [ - G_l G'_k \A_{likj}(\br)  - \frac{i}{2} (G_m + G'_m) \B^-_{mij}(\br)] \psi_{jp} (\bG', \bk) \nonumber \\
    &= \sum_{\bG'} \int \frac{d^2 \br}{A_M} e^{i \bG \cdot \br} [-i G_l \A_{likj}(\br) \partial_k  - \frac{i}{2} \B^-_{mij}(\br) (G_m + i \partial_m) ] e^{-i \bG' \cdot \br} \psi_{jp} (\bG', \bk) \nonumber \\
    &= \int \frac{d^2 \br}{A_M} \sum_{\bG'}   e^{-i \bG' \cdot \br} \psi_{jp} (\bG', \bk) e^{i \bG \cdot \br} [i G_l \partial_k \A_{likj}(\br) - G_l G_k \A_{likj}(\br)   - i G_m \B^-_{mij}(\br) - \frac{1}{2} \partial_m \B^-_{mij}(\br)]   \nonumber \\
    &= \int \frac{d^2 \br}{A_M} e^{i \bG \cdot \br} \psi_{jp} (-\br, \bk)  [i G_l \partial_k \A_{likj}(\br) - G_l G_k \A_{likj}(\br)   - i G_m \B^-_{mij}(\br)] \nonumber \\
    &= \int \frac{d^2 \br}{A_M} e^{i \bG \cdot \br}  \partial_m [ \A_{mikj}(\br) \partial_k  \psi_{jp} (-\br, \bk) +  \B^-_{mij}(\br) \psi_{jp} (-\br, \bk)] .
\end{align}
\end{widetext}

On the other hand, we can use the direct expression for $\Gamma_{ip}$ [Eq.~\eqref{GammaDef}]
\begin{align}
    &\Gamma_{ip}(\bG + \bk, \bk) \nonumber\\
    &= -\int \frac{d^2 \br}{A_M} e^{i \bG \cdot \br} [G_l k_k \A_{likp}(\br) + i k_m \B^-_{mip}(\br)] \nonumber \\
    &= -i k_m \int \frac{d^2 \br}{A_M} e^{i \bG \cdot \br} [ \partial_l \A_{limp}(\br) + \B^-_{mip}(\br)].
\end{align}
Thus $\psi_{jp}(-\br, \bk)$ solves the equation
\begin{multline}
    [\partial_l \A_{limj}(\br) + \B^-_{mij}(\br)] \partial_m  \psi_{jp} (-\br, \bk) + \\ \A_{mikj}(\br) \partial_m \partial_k  \psi_{jp} (-\br, \bk)  = -i k_m  [ \partial_l \A_{limp}(\br) + \B^-_{mip}(\br)].
\end{multline}
It is obvious from this equation that $\psi_{jp}(\br,\bk)$ is a linear function of $\bk$ so we can write it as $\psi_{jp}(\br,\bk) = i k_l\psi_{jpl}(\br)$ which satisfies the equation
\begin{multline}
    [ \partial_k \A_{kimj}(\br) +  \B^-_{mij}(\br)] \partial_m  \psi_{jpl} (-\br) + \\ \A_{mikj}(\br) \partial_m \partial_k  \psi_{jpl} (-\br)  = -[ \partial_m \A_{milp}(\br) + \B^-_{lip}(\br)].
\end{multline}
This equation can be solved in the vicinity of a domain wall by replacing $\A$ and $\B$ by $\A^{\rm DW}$ and $\B^{\rm DW}$ from Eq.~\eqref{DWMatrices}. We first note that $\B^{\rm DW}$ vanishes upon antisymmetrizing relative to the two last indices so that $\B^-$ drops out. Using Eq.~\eqref{DWMatrices}, we can see that we can choose all entries of $\psi^{\rm DW}_{jpl}$ to vanish except $jpl = xxx$. The non-zero entry satisfies
\begin{equation}
\begin{split}
     &\mu g'(x)^2 \partial^2_x  \psi^{\rm DW}_{xxx} (-\br) + \mu (\partial_x g'(x)^2) \partial_x  \psi^{\rm DW}_{xxx} (-\br)\\
     &= -\mu \partial_x g'(x)^2,
\end{split}
\end{equation}
which can be solved by choosing $\psi_{xxx}(-\br) = -x$. Substituting in Eq.~\eqref{DeltaGammaPsi}, we get
\begin{equation}
\begin{split}
    \Delta \Gamma^{\rm DW}_{xx} &= -k_x^2 \int \frac{d^2 \br}{A_M}  \psi^{\rm DW}_{xxx}(-\br, \bk) \mu \partial_x \phi_y'(x)^2 \\
    &= -\mu k_x^2 \langle g'(x)^2 \rangle,
\end{split}
\end{equation}
Thus, the result only depends on $\langle g'(x)^2 \rangle$ which can be easily evaluated as
\begin{align}
    \beta &= \langle g'(x)^2 \rangle = \frac{3}{(2\pi)^2} \int \frac{d^2 \br}{A_M} f'(x)^2 \nonumber \\
    &= \frac{\sqrt{3} \theta a}{2 \pi^2 w^2} \int_{-\infty}^\infty d x \frac{9}{2 (1 + 2 \cosh \frac{ \sqrt{3}}{2 w} x)^2} \nonumber \\
    &= \frac{\sqrt{3} \theta a}{2 \pi^2 w} (\sqrt{6} - \frac{\sqrt{2}}{3}\pi).
\end{align}
Substituting in Eq.~\eqref{DeltaGamma}) and using Eq.~\eqref{DWMatrices} gives
\begin{align}
    \Gamma^{\rm DW}_{ij} &= \delta_{i,x} \delta_{i,y} \beta [\omega^2 \rho  - k_x^2 \mu - k_y^2 (\lambda + 2\mu) + \mu k_x^2] \nonumber \\&=  \delta_{i,x} \delta_{i,y} \beta [\omega^2 \rho - k_y^2 (\lambda + 2\mu)].
\end{align}
Symmetrizing with respect to the three domain wall directions yields
\begin{equation}
    \Gamma = \frac{3}{2} \beta \rho \omega^2 \left(\begin{array}{cc}
        1 & 0 \\
        0 & 1
    \end{array} \right) - \frac{3}{8} \beta (\lambda + 2\mu) \left(\begin{array}{cc}
        k_x^2 + 3 k_y^2 & -2 k_x k_y \\ -2 k_x k_y & 3 k_x^2 + k_y^2
    \end{array} \right)
\end{equation}
from which we can read off
\begin{gather}
    \tilde \rho = 3 \beta \rho, \qquad \tilde \lambda = -\frac{3}{4} \beta (\lambda + 2 \mu), \nonumber \\ \tilde \mu = \frac{3}{4} \beta (\lambda + 2 \mu), \qquad \tilde \gamma = \frac{3}{2} \beta (\lambda + 2 \mu).
\end{gather}

\section{Third-order anharmonic term}\label{thirdOrder}
In this appendix, we show explicitly that the coupling between the soft mode and other modes due to third order anharmonic term in the Lagrangian vanishes at least linearly in the momentum $\bq$ of the soft mode.

Since the relation between the variable $\bxi$ and $\delta \bu$ [Eq.~\eqref{NonLinearTransformation}] is generally non-linear, to expand the Lagrangian to the third order, we need to write $\delta \bu$ in terms of $\bxi$ up to quadratic order:
\begin{multline}
    \delta u_i = \xi_j\partial_j u_i + (1- \mathcal{R}^{-1}_{\theta})_{ij}\xi_j + \frac{1}{2}\xi_j\xi_k\partial_j\partial_k u_i \\ \equiv \mathcal{C}_{ij}\xi_j+ \frac{1}{2}\xi_j\xi_k\partial_j\partial_k u_i.
\end{multline}
Substituting in the Lagrangian and expanding to third order in $\bxi$ yields $\mathcal{L}^{\text{3rd}} = \mathcal{T}^{\text{3rd}} - \mathcal{V}^{\text{3rd}}$ where 
\begin{gather}
   \mathcal{T}^{\text{3rd}} = \frac{\rho}{2}\mathcal{C}_{li}\partial_m\partial_j u_l\xi_m\dot{\xi}_i\dot{\xi}_j, \\ \mathcal{V}^{\text{3rd}} =  \frac{1}{2}\bigg[\mathcal{E}_{ijklm}\xi_i\partial_l\xi_j\partial_m\xi_k  + \mathcal{F}_{ijkl}\xi_i\xi_j\partial_l\xi_k  + \mathcal{G}_{ijk}\xi_i\xi_j\xi_k\bigg].
\end{gather}
The explicit expressions for $\mathcal E$, $\mathcal F$, and $\mathcal G$ are periodic functions of $\br$.

We note that the symmetry acts on $\bxi$ as $\bxi \mapsto \bxi + \bepsilon$ where $\bepsilon$ is a constant. This implies that $\frac{d}{d \bepsilon} \mathcal{V}^{\text{3rd}}[\bxi + \bepsilon]|_{\bepsilon = 0} = \frac{d}{d \bxi} \mathcal{V}^{\text{3rd}}[\bxi]$ is a total derivative:
\begin{equation}
    \mathcal{E}_{ijklm} \partial_l\xi_j\partial_m\xi_k + 2 \mathcal{F}_{ijkl}\xi_j\partial_l\xi_k  + 3\mathcal{G}_{ijk} \xi_j\xi_k = \partial_r \H_{ir}
    \label{Sym3}
\end{equation}
for some arbitrary function $\H_{ir}$. Writing $\int d^2 \br \mathcal V$ in momentum space as
\begin{multline}
     \int d^2 \br \mathcal V = \frac{1}{2}\sum_{\bq_1, \bq_2, \bq_3, \bG} \bigg\{-q_{2l} q_{3m} \mathcal{E}_{ijklm}(\bG)  + i q_{3l} \mathcal{F}_{ijkl}(\bG) \\  + \mathcal{G}_{ijk}(\bG)\bigg\}  \xi_i(\bq_1)  \xi_j(\bq_2)  \xi_k(\bq_2) \delta_{\bq_1 + \bq_2 + \bq_3 + \bG,0}.
\end{multline}
We want to consider the coupling to the soft mode whose momentum $\bq = \bdelta \rightarrow 0$ and show that the coupling goes as $\bdelta$. For the first term, this will be the case if $\bq_2$ or $\bq_3$ is $\bdelta$, so we  only need to consider the case $\bq_1 = \bdelta$. For the second term, we need to consider both terms with $\bq_1 = \bdelta$ or $\bq_2 = \bdelta$ which are equal by symmetry yielding a factor of 2. Similarly, for the third term, we get three contributions if any of $\bq_1$, $\bq_2$ or $\bq_3$ is equal to $\bdelta$ yielding a factor of 3. Thus,
\begin{multline}
    \int d^2 \br \mathcal V = \frac{1}{2}\sum_{\bq_2, \bq_3, \bG} \bigg\{-q_{2l} q_{3m} \mathcal{E}_{ijklm}(\bG) + 2 i q_{3l} \mathcal{F}_{ijkl}(\bG) \\ + 3 \mathcal{G}_{ijk}(\bG)\bigg\} \xi_i(\bdelta)  \xi_j(\bq_2)  \xi_k(\bq_2) \delta_{\bdelta + \bq_2 + \bq_3 + \bG,0} \\ = -\frac{1}{2} \delta_r \H_{ir}(-\bdelta) \xi_i(\bdelta) \rightarrow 0, \qquad \qquad \quad
\end{multline}
where we used Eq.~\eqref{Sym3}. We emphasize here that this asymptotic behavior requires the function $\H$ to have no poles at $\bdelta= 0$, which then requires that the coefficients $\mathcal E$, $\mathcal F$, and $\mathcal G$ in $\mathcal{V}^{\text{3rd}}$ also have no poles at $\bdelta= 0$. This is true in TBG for any finite twisting angles, where the displacement $\bu$ is smooth and so are the coefficients since they can be expressed in terms of the spatial derivatives of $\bu$.

\section{Rotation Adapted Mirror Symmetry in Alternating Twisted Multilayer Graphene}\label{symmATMG}
The alternating twisted multilayer graphene (ATMG) including TBG has a geometrical symmetry that the system is invariant under a $C_2$ rotation around an in-plane axis. If we are interested in the in-plane motion of the atoms, then such symmetry can be projected onto the plane as effective mirror symmetry. However, the binding energy written in the main text (for example, the Eq.~\eqref{bindingEnenrgy}) does not preserve the mirror symmetry. Instead, the binding energy has a rotation adapted mirror symmetry. We want to illustrate this aspect by using the geometry of TBG as an example and then extend it to a more general scenario.

For TBG, the equilibrium equations for the displacement fields of two layers, $\bu^{(1)}(\br)$ and $\bu^{(2)}(\br) $, read
\begin{equation}
    \begin{split}
        (\lambda + \mu)\boldsymbol{\nabla}(\boldsymbol{\nabla}\cdot\bu^{(1)}) + \mu\boldsymbol{\nabla}^2\bu^{(1)} &= \frac{\partial V^{(21)}}{\partial \bu^{(1)}}, \\
        (\lambda + \mu)\boldsymbol{\nabla}(\boldsymbol{\nabla}\cdot\bu^{(2)}) + \mu\boldsymbol{\nabla}^2\bu^{(2)} &= \frac{\partial V^{(21)}}{\partial \bu^{(2)}},
    \end{split}
\end{equation}
where
\begin{equation}
    V^{(21)} = 2V_0\sum_{i=1}^3 \cos[\boldsymbol{G}^{(12)}_i\cdot\boldsymbol{r} + \boldsymbol{a}^{*(1)}_i \cdot (\bu^{(2)}(\boldsymbol{r})-\bu^{(1)}(\boldsymbol{r}))].
\end{equation}
Here $ \boldsymbol{G}^{(lm)}_i = \boldsymbol{a}^{*(l)}_i-\boldsymbol{a}^{*(m)}_i$ is the moir\'e reciprocal lattice vector between the $l^{\text{th}}$ and $m^{\text{th}}$ layers. We have utilized the fact that $\ba^{*(1)}_i \cdot(1 - \mathcal{R}_{\theta}^{-1})\br = \boldsymbol{G}^{(12)}_i\cdot\boldsymbol{r}$.
Now, consider a mirror action along one of the mirror-symmetric axes of the Mori\'e Brillouin zone, and without loss of generality, let it be the one parallel to $\boldsymbol{G}^{(12)}_1$. Such action can be denoted as: $\br\to\br' = \sigma_{v_1}\br $. Under this action, we then have
\begin{equation}
\begin{split}
    \boldsymbol{G}^{(12)}_1\cdot\br' &= \boldsymbol{G}^{(12)}_1\cdot\br, \qquad \boldsymbol{G}^{(12)}_2\cdot\br' = \boldsymbol{G}^{(12)}_3\cdot\br, \\ \boldsymbol{G}^{(12)}_3\cdot\br' &= \boldsymbol{G}^{(12)}_2\cdot\br.
\end{split}
\end{equation}
For the equilibrium equation to be invariant under such transformations, we require that the displacement fields transform accordingly. It is not hard to see that if
\begin{equation}
\begin{split}
    \ba^{*(1)}_1\cdot \bu^{(1)}(\br') &= -  \ba^{*(1)}_1\cdot \bu^{(2)}(\br), \\
    \ba^{*(1)}_2\cdot \bu^{(1)}(\br') &= -  \ba^{*(1)}_3\cdot \bu^{(2)}(\br), \\
    \ba^{*(1)}_3\cdot \bu^{(1)}(\br') &= -  \ba^{*(1)}_2\cdot \bu^{(2)}(\br),
\end{split}
\end{equation}
and similarly
\begin{equation}
\begin{split}
    \ba^{*(1)}_1\cdot \bu^{(2)}(\br') &= -  \ba^{*(1)}_1\cdot \bu^{(1)}(\br), \\
    \ba^{*(1)}_2\cdot \bu^{(2)}(\br') &= -  \ba^{*(1)}_3\cdot \bu^{(1)}(\br), \\
    \ba^{*(1)}_3\cdot \bu^{(2)}(\br') &= -  \ba^{*(1)}_2\cdot \bu^{(1)}(\br),
\end{split}
\end{equation}
the equilibrium equation is invariant. In a more symbolic representation, we can write
\begin{equation}\label{n2case}
\begin{split}
    \bu^{(2)}(\br) &= \mathcal{R}^{-1}_{\theta}\sigma_{v_1}\bu^{(1)}(\sigma_{v_1}\br), \\
    \bu^{(1)}(\br) &= \sigma_{v_1}\mathcal{R}_{\theta}\bu^{(2)}(\sigma_{v_1}\br),
\end{split}
\end{equation}
where the rotation enters the argument. Then the relative displacement field $\bu^{(21)}(\br)$ inherits symmetry:
\begin{equation}
    \bu^{(21)}(\br) = -\sigma_{v_1}\mathcal{R}_{\theta}\bu^{(21)}(\sigma_{v_1}\br),
\end{equation}
where we have used the fact that $\mathcal{R}^{-1}_{\theta}\sigma_{v_1}=\sigma_{v_1}\mathcal{R}_{\theta} $.

Similar discussion can be applied to ATMG with more $n>3$ layers. However, one should keep in mind that the mirror symmetry only relates the $l$-th layer to the $(n-l+1)$-th layer, not the adjacent two layers. The symmetry has the following form:
\begin{equation}
    \bu^{(l)}(\boldsymbol{r})= (\sigma_v \mathcal{R}_{(-1)^{l-1}\theta})^{n+1}\bu^{(n+1-l)}(\sigma_v^{n+1}\boldsymbol{r}), \text{for } l \leq n/2.   
\end{equation}
We note that for ATMG with an even number of layers, the symmetry is geometrically the same as in TBG which is a $C_2$ rotation symmetry around an in-plane axis, while for ATMG with an odd number of layers, the symmetry becomes an exact mirror symmetry with the middle layer as the mirror plane.
For example, if $n=2$, we have the symmetry described by Eq.~\eqref{n2case}; if $n=3$, we have $\bu^{(1)}(\boldsymbol{r}) = \bu^{(3)}(\boldsymbol{r})$. An important observation is that the symmetry is non-local for $n$ being even while becomes local for $n$ being odd. Finally, we want to emphasize here that the symmetry requires an ``AA" stacking of layers when $\theta = 0$.


%

\end{document}